\documentclass[
  reprint,
  prl,
  aps,
  superscriptaddress
]{revtex4-2}
\usepackage{amsmath,amssymb}
\usepackage{graphicx}
\usepackage{bm}
\usepackage{physics}
\usepackage{xcolor}
\usepackage[squaren, Gray, cdot]{SIunits}
\usepackage{hyperref}
\usepackage{comment}
\usepackage{caption}
\usepackage{cancel}
\usepackage{tikz,pgfplots}
\usepackage{pgfplotstable}
\usepackage[normalem]{ulem}
\usepgfplotslibrary{groupplots}
\usepackage{booktabs}
\pgfplotsset{compat=1.18}

\hypersetup{
    colorlinks = true,
    linkcolor = blue,
    anchorcolor = blue,
    citecolor = red,
    filecolor = blue,
    urlcolor = blue
    }

\newcommand{\m}[1]{\begin{pmatrix} #1 \end{pmatrix}}

\newcommand{\defeq}{\stackrel{\mathrm{def}}{=}}
\definecolor{darkgreen}{RGB}{0,100,0}

\newcommand{\ii}{\mathrm{i}}
\newcommand{\ff}{\mathrm{f}}

\begin{document}

\title{Controlling transitions between nonequilibrium states through active bath engineering}

\author{Dima Boriskovsky}
\email{boriskovsky@unistra.fr}
\affiliation{
    Institut de Sciences et d'Ingénierie Supramoléculaires, University of Strasbourg, France
}
\author{Salamb\^{o} Dago}
\email{sdago@unistra.fr}
\affiliation{
    Institut de Sciences et d'Ingénierie Supramoléculaires, University of Strasbourg, France
}
\author{Cyriaque Genet}
\email{genet@unistra.fr}
\affiliation{
    Institut de Sciences et d'Ingénierie Supramoléculaires, University of Strasbourg, France
}
\author{R\'{e}mi Goerlich}
\altaffiliation[Present address:]{
    Laboratoire de Physique de l'ENS de Lyon, CNRS, 
    ENS de Lyon, Lyon, France
}
\affiliation{
    Institut de Sciences et d'Ingénierie Supramoléculaires, University of Strasbourg, France
}
\author{Vincent Hardel}
\affiliation{%
    Institut de Physique et Chimie des Mat\'eriaux de Strasbourg, France
}%
\author{Paul-Antoine Hervieux}
\email{paul-antoine.hervieux@ipcms.unistra.fr}
\affiliation{%
    Institut de Physique et Chimie des Mat\'eriaux de Strasbourg, France
}%
\author{Harry L.F. Ip}
\email{harryip2-c@my.cityu.edu.hk}
\affiliation{%
    City University of Hong Kong, Hong Kong, China
}%
\author{Giovanni Manfredi}
\email{giovanni.manfredi@ipcms.unistra.fr}
\affiliation{%
    Institut de Physique et Chimie des Mat\'eriaux de Strasbourg, France
}%
\author{Laurent Mertz}
\email{lmertz@cityu.edu.hk}
\affiliation{%
    City University of Hong Kong, Hong Kong, China
}%

\date{\today}

\begin{abstract}
We develop a family of control protocols for finite-time transitions between active nonequilibrium steady states using the noise-color (correlation rate) as the sole control parameter. 
By reverse-engineering the second moment dynamics of an active Ornstein-Uhlenbeck process, we determine the time-dependent correlation rate required to realize a prescribed evolution of the system's state. 
We experimentally implement these protocols with a micrometer-sized optically trapped particle coupled to an engineered active bath, demonstrating transitions substantially faster than the natural relaxation while keeping the confining potential, noise amplitude, and temperature fixed. 
Our approach extends engineered swift-equilibration methods to far-from-equilibrium steady states by directly controlling the temporal correlations of an active environment. 
We show how physical constraints impose a speed limit on finite-time transitions, while the freedom in choosing the prescribed system evolution can be exploited to eliminate control discontinuities, minimize the admissible transition time, or optimize a thermodynamic cost.
\end{abstract}

\maketitle

{\it Introduction}.---
Accelerating and controlling transitions between steady states of physical systems is a central challenge in statistical physics and thermodynamics, with potential applications to micro- and nanoscale engines \cite{Seifert2012,Pietzonka2018,Krishnamurthy2023}, information processing \cite{Zulkowski2014,Proesmans2020,Giorgini2023,Barros2025}, and active matter \cite{Davis2024,Alvarado2026}. 
While most finite-time control strategies have been developed for thermodynamic transformations between equilibrium states, many natural and engineered systems operate far from equilibrium, motivating their extension to nonequilibrium steady states (NESS). 

Active systems provide a paradigmatic class of NESS that operate far from equilibrium, with characteristic phenomena such as self-propulsion and emergent collective behavior \cite{Ramaswamy2010,Cates2015,Fodor2016}. 
Whether activity is intrinsic to the particles or imparted by an active environment, these states are sustained by continuous energy injection and dissipation, and generally violate the fluctuation-dissipation theorem. 
Controlling active systems is essential for understanding and modeling self-organized matter and collective motion, including cytoskeletal restructuring \cite{Lamtyugina2022,Ideses2013,SonnSegev2017}, cell rounding \cite{Taubenberger2020,Hurst2021}, transport in Brownian ratchets \cite{Paneru2021,Ghosh2013}, and microswimmers \cite{Bechinger2016}.

For passive Brownian systems, swift-equilibration strategies are well established for both isothermal and isochoric processes \cite{Martinez2016,LeCunuder2016,Chupeau2018,Chupeau2018_BE,RosalesCabara2020,Pires2023,Rengifo2025}. 
By controlling the confining potential or thermal noise, these methods drive transitions between equilibrium states faster than the system's natural relaxation time. 
Such protocols have been developed for both overdamped \cite{Martinez2016} and underdamped \cite{Chupeau2018} regimes, including optimal-control strategies that minimize thermodynamic costs such as work or entropy production \cite{Schmiedl2007,RosalesCabara2020,Patron2022,Pires2023,Oikawa2025,Cocconi2026}.

For active systems, analogous control strategies remain less developed and are the subject of ongoing research.
Recent approaches address active Brownian particles \cite{Baldovin2023,Olsen2025}, active Ornstein-Uhlenbeck particles (AOUPs), and run-and-tumble particles \cite{Frim2023,Baldovin2026}. 
For AOUPs in particular, protocols controlling the confining potential have been designed using linear-response theory \cite{Gupta2023,Davis2024}, neural networks \cite{Casert2024}, and closed-loop feedback \cite{Garcia-Millan2025}.

An emerging alternative is to control the active fluctuations themselves, through either their amplitude \cite{Olsen2026} or temporal correlations \cite{Goerlich2022b}. 
The latter is particularly intriguing as a genuine property of a nonequilibrium bath: for a passive particle in an active environment, the \emph{color} of the active noise can serve as the sole control parameter.
To our knowledge, no finite-time protocols on the persistence time of AOUPs exist in theory or experiment.

Here, we develop a family of control protocols based on arbitrary modulation of the active correlation rate $\omega_c(t)$ (the inverse persistence time).
These protocols control NESS-to-NESS transition times, allowing acceleration or deceleration relative to the natural relaxation time while keeping the trap stiffness, temperature, and noise amplitude fixed. 
We experimentally implement these protocols using a trapped colloidal particle in an engineered active bath and find excellent agreement with theory. 

{\it Model}.---
We consider an AOUP confined in a harmonic trap subject to thermal fluctuations
\begin{equation}
\label{eq:x_process}
    \dot{x}(t) = -\omega_0\, x(t) + \sqrt{2D}\, \xi_x(t) + \sqrt{2D_a}\, \eta(t),
\end{equation}
where the active noise $\eta(t)$ evolves according to the ancillary Ornstein--Uhlenbeck (OU) dynamics
\begin{equation}
\label{eq:eta_process}
    \dot{\eta}(t) = -\omega_c(t)\, \eta(t) + \sqrt{2\alpha \omega_c(t)}\, \xi_\eta(t).
\end{equation}
Here, $\xi_x(t)$ and $\xi_\eta(t)$ are independent Gaussian white noises with zero mean and unit variance. The trap relaxation rate
$\omega_0 = \kappa/\gamma$
is the ratio of $\kappa$, the stiffness of the confining potential 
$U(x) = \tfrac12 \kappa x^2$, to $\gamma$, the viscous drag coefficient.  
$D_a$ is the active diffusion coefficient.
The thermal diffusion coefficient $D = k_B T/\gamma$ ensures that the fluctuation--dissipation theorem holds when $D_a = 0$. 
The correlation rate $\omega_c(t)$ is used as a time-dependent control variable, while all other parameters are kept fixed.

{\it Finite-time  transitions via colored-noise protocols}.---
Consider the second moments:
variance,
$s(t) \equiv \langle x^2(t)\rangle$, and covariance,
$z(t) \equiv \langle x(t)\eta(t)\rangle$.
Using the AOUP dynamics \eqref{eq:x_process}-\eqref{eq:eta_process}, these moments satisfy   
\begin{align}
\dot{s}(t) &= -2\omega_0\,s(t) + 2C\,z(t) + 2D,
\label{eq:s_dyn}
\\[2mm]
\dot{z}(t) &= -(\omega_0+\omega_c(t))\,z(t) + C\alpha,
\label{eq:z_dyn}
\end{align}
where $C=\sqrt{2D_a}$.  
Note that $\langle \eta^2(t)\rangle=\alpha$ for all $t>0$, provided that the initial condition satisfies $\langle \eta^2(0)\rangle=\alpha$.

Our reverse-engineering approach relies on first expressing $z$ in terms of $s$ and $\dot{s}$, namely,
\(
z(t)=(-2D+2\omega_0 s(t)+\dot{s}(t))/2C.
\)
Differentiating this expression and substituting Eq. \eqref{eq:z_dyn} yields an explicit inversion formula for the control parameter in terms of the variance of $x$ and its derivatives up to second order:
\begin{equation}
\label{eq:omegac}
\omega_c(t)
=
\frac{2C^2\alpha-\bigl[2\omega_0\dot{s}(t)+\ddot{s}(t)\bigr]}
{-2D+2\omega_0 s(t)+\dot{s}(t)}
-\omega_0.
\end{equation}
This formula constitutes the core of our method.

Although we focus here on the simple OU case, our results are applicable to a wide variety of stationary Gaussian noises; see Sec.~VII of Supplemental Material (SM) \cite{SM}.
They open new avenues for the experimental engineering of NESSs and for the development of a mathematical theory of controllability in active systems.

From a control-theoretic perspective, Eq.~\eqref{eq:omegac} provides an explicit controllability map between admissible variance trajectories and realizable noise-color control protocols. 
The finite-time NESS control problem is thus reduced to the construction of an admissible variance trajectory $s(t)$. Once $s(t)$ is specified, both the covariance $z(t)$ and the required noise-color protocol $\omega_c(t)$ follow uniquely and exactly.

Furthermore, one must impose $z(t)>0$ to ensure the denominator of Eq. \eqref{eq:omegac} never vanishes and $\omega_c(t)>0$ to guarantee a positive correlation time along with the physical admissibility of the reconstructed moments (see Sec.~I of the SM \cite{SM}).
The control problem is therefore reduced from the two-dimensional stochastic dynamics $(x,\eta)$ to the construction of a single scalar function $s(t)$. Polynomial choices of $s(t)$ will provide explicit experimentally realizable protocols.

\begin{figure}[t]
    \centering
    \includegraphics[width=0.45\textwidth]{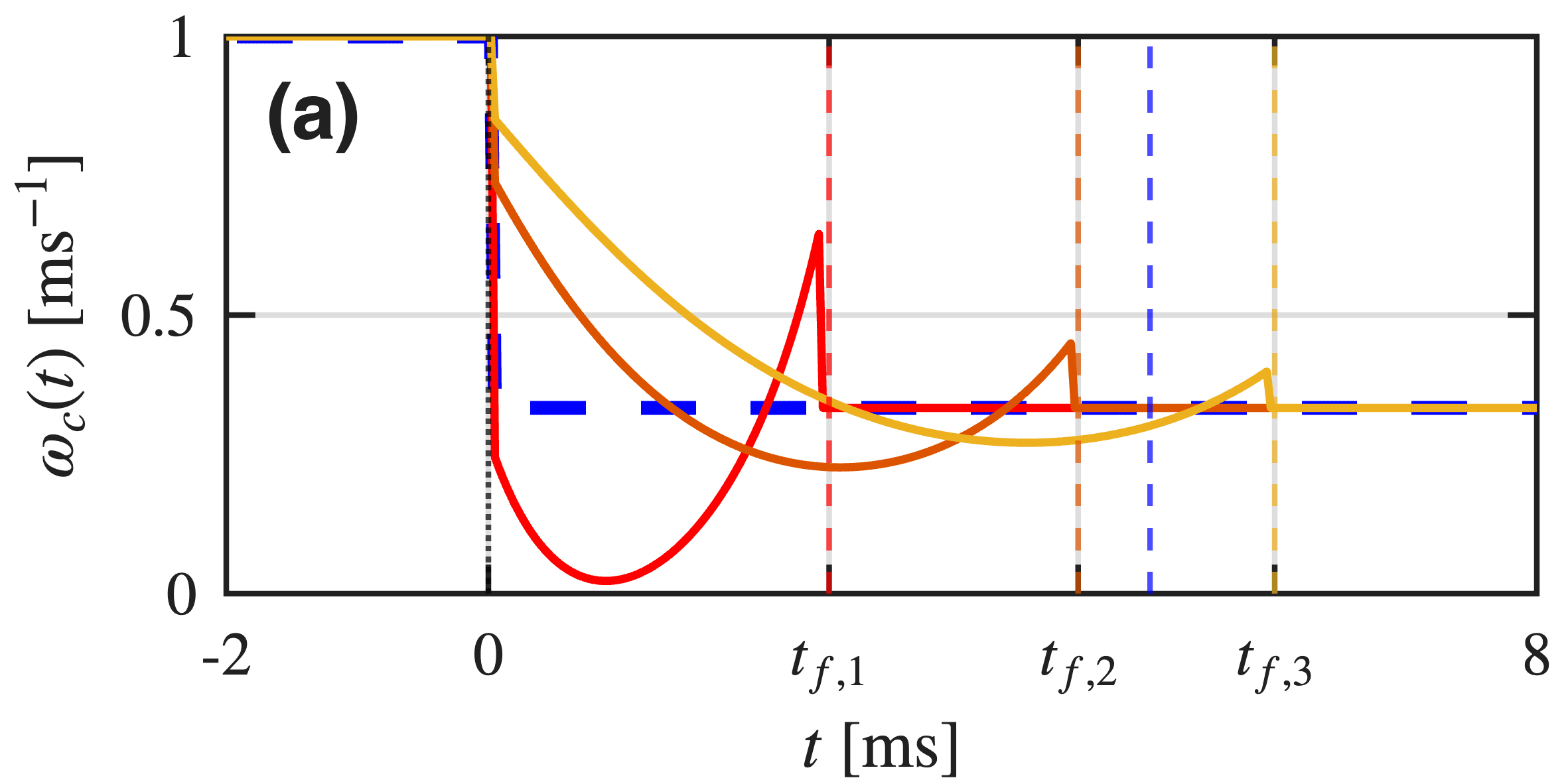}
    \includegraphics[width=0.45\textwidth]{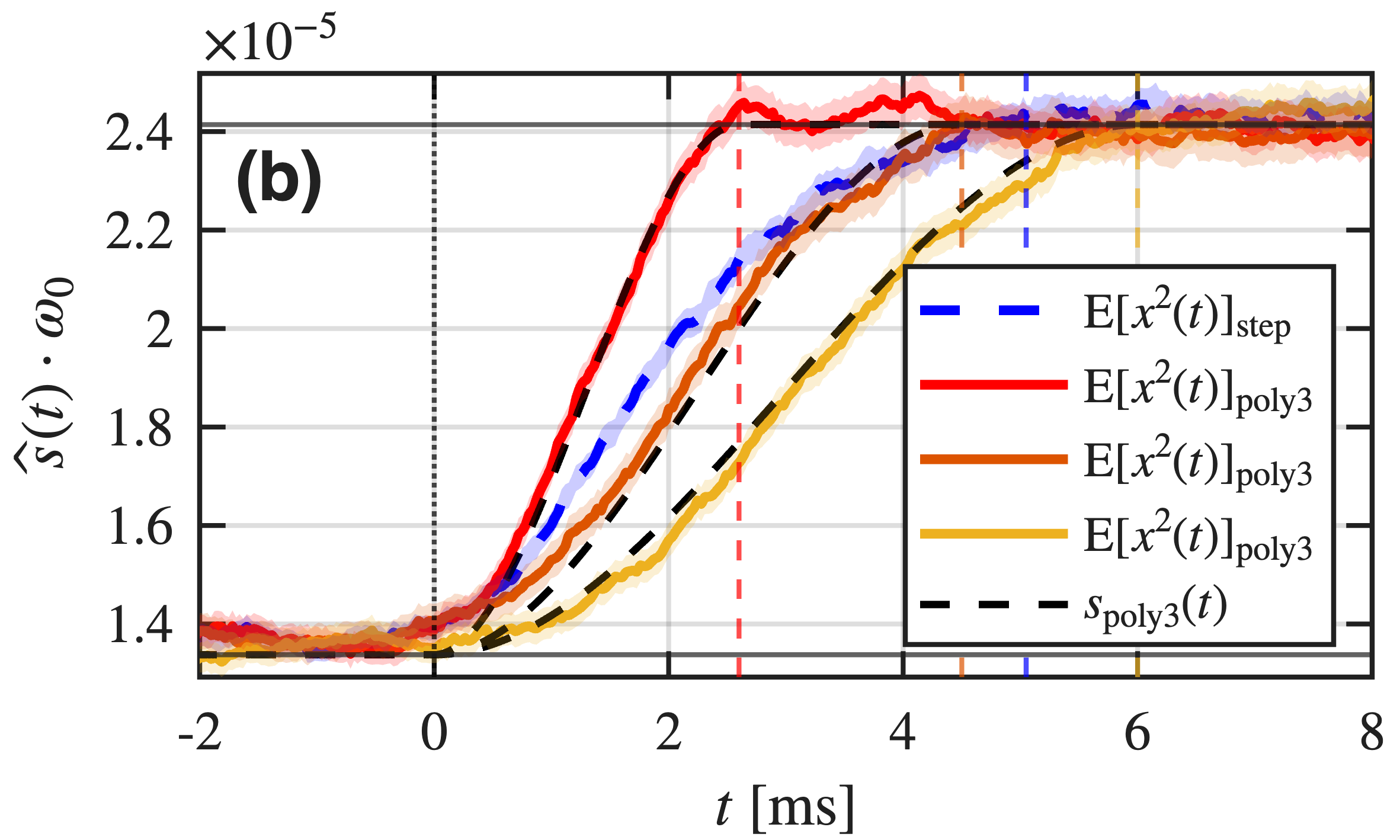}
    \captionsetup{justification=raggedright,singlelinecheck=false}
    \caption{
    NESS-to-NESS transition control.
    \textbf{(a)} Third-order polynomial control protocols $\omega_c(t)$ for expansion with transition times $t_f=2.60$, $4.50$, and $6~\mathrm{ms}$, compared with an instantaneous STEP protocol (blue dashed curve).
    \textbf{(b)} Corresponding experimental normalized variances $\hat{s}(t)$ together with the analytical predictions of Eq.~\eqref{eq:poly3} (black dashed curves). 
    Vertical dashed lines mark the prescribed $t_f$, shaded regions indicate standard errors, and horizontal lines denote the NESS values $\hat{s}_{i,f}$ (Eq.~\eqref{eq:NESS}). 
    Experimental parameters are $D=1.41\times10^{-4}~\mu\mathrm{m}^2/\mathrm{ms}$, $\omega_0=0.49~\mathrm{ms}^{-1}$, $\alpha=0.01~\mathrm{Hz}$, and $\omega_c$ is varied between $1$ and $\frac{1}{3}~\mathrm{ms}^{-1}$. 
    Ensemble averages are obtained from $M\approx10^4$ trajectories over a $40$-ms observation window (not fully displayed here).
    Further details of the experimental procedure and data analysis are provided in Sec.~VI of SM \cite{SM}.
    }
    \label{fig:fig2}
\end{figure}

We consider a transition between two NESS associated with correlation rates $\omega_{c,i}$ and $\omega_{c,f}$. 
Since the dynamics is Gaussian with zero mean, the state of the system is fully characterized by the second moments $s(t)$ and $z(t)$.
At steady state, the variances are
\begin{equation}
\label{eq:NESS}
s_{i,f}
=
\frac{D}{\omega_0}
+
\frac{2D_a\alpha}
{\omega_0\bigl(\omega_0+\omega_{c,i,f}\bigr)}.
\end{equation}
Since the system must be stationary before and after the protocol,
continuity of the moments together with the moment equations imply
\begin{equation}
\label{eq:sif}
s(0)=s_i,\; s(t_f)=s_f,\quad \dot s(0)=\dot s(t_f)=0,
\end{equation}
where $t_f$ is the duration of the transition. These conditions likewise fix $z(t)$ to its steady-state values at the boundaries, while still allowing finite discontinuous jumps of $\omega_c(t)$ (see Sec.~I of the SM \cite{SM}).

\begin{figure*}[t]
    \centering
    \includegraphics[width=0.45\textwidth]{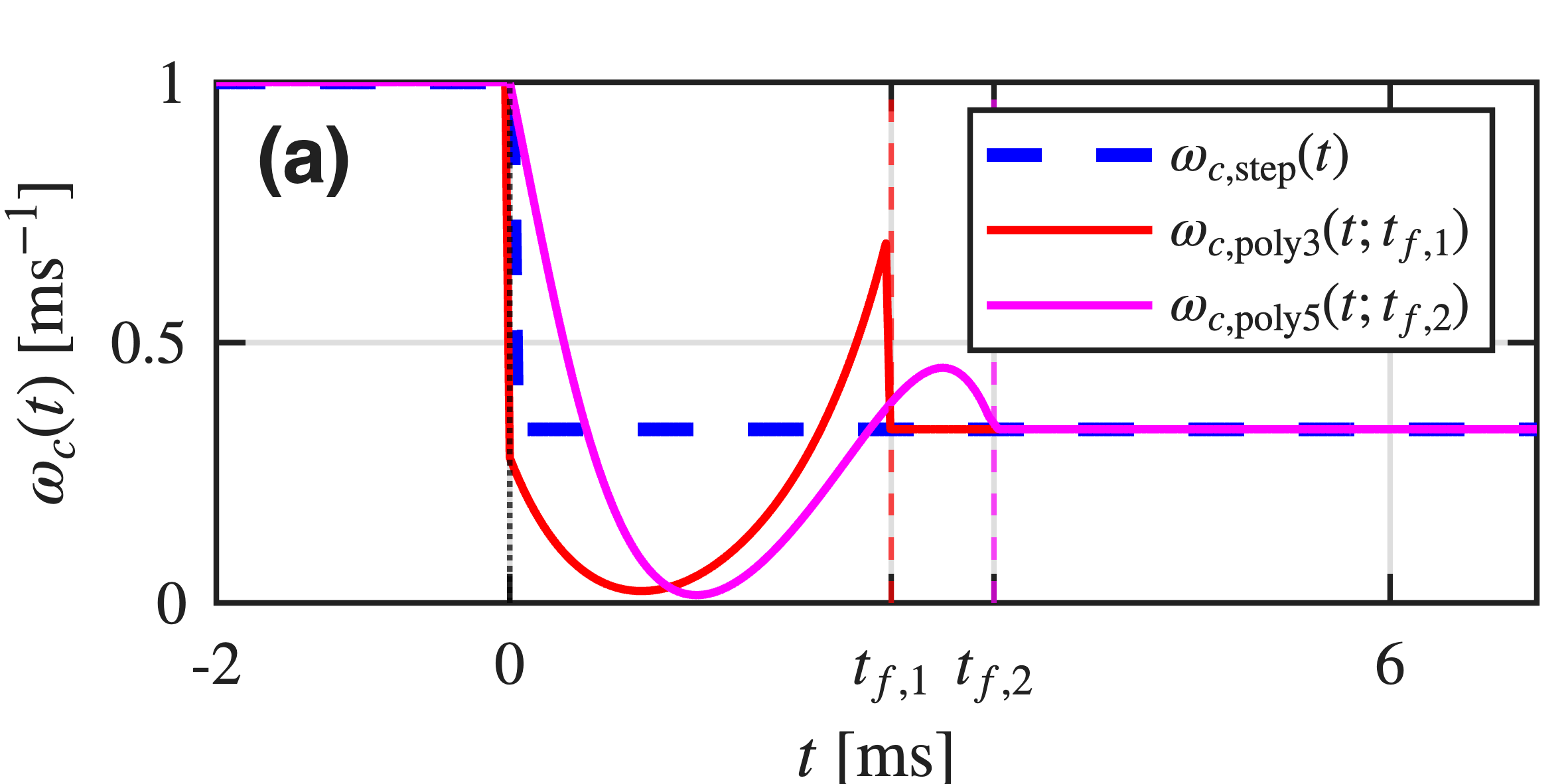} 
    \includegraphics[width=0.45\textwidth]{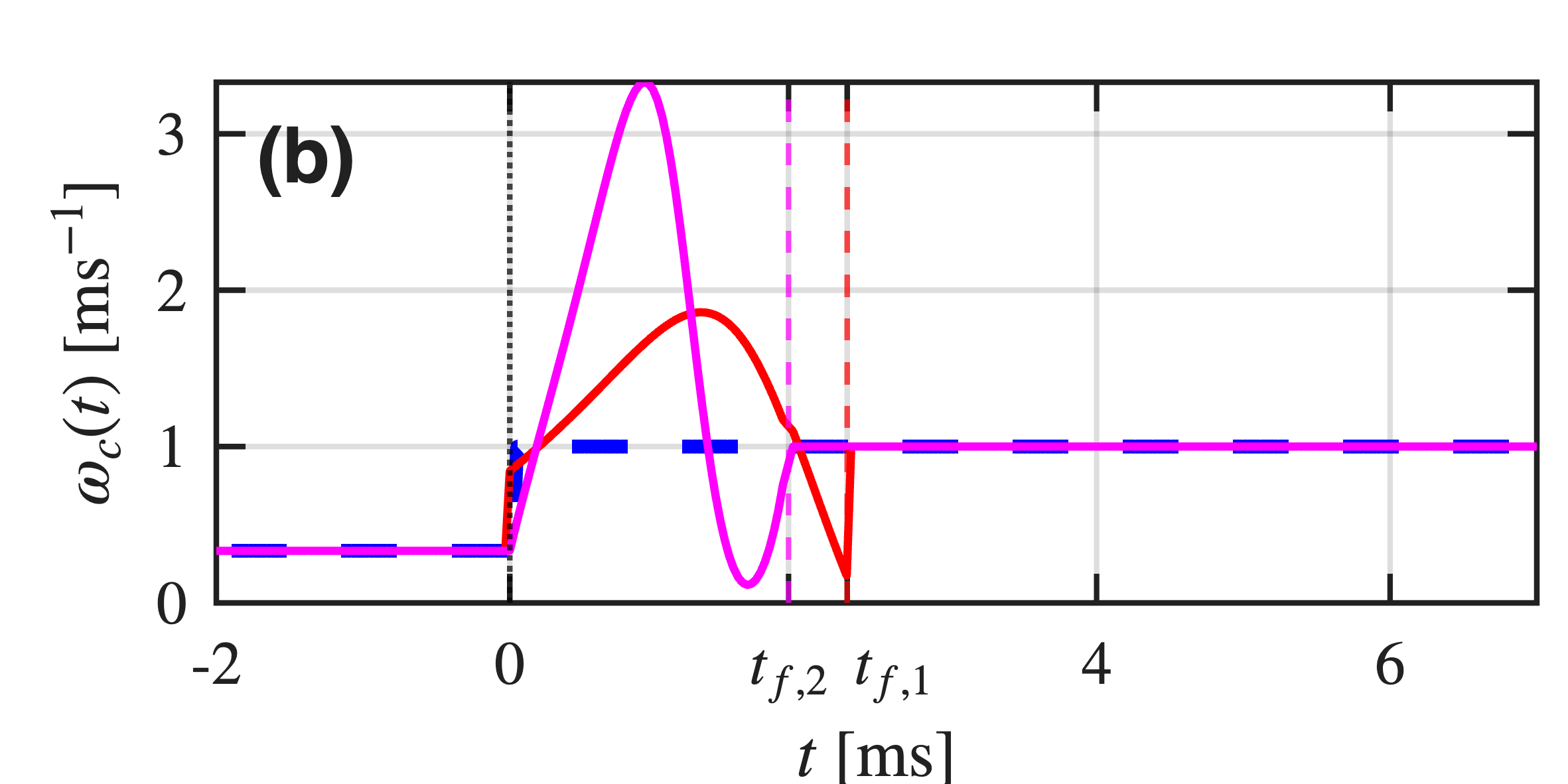}

    \includegraphics[width=0.45\textwidth]{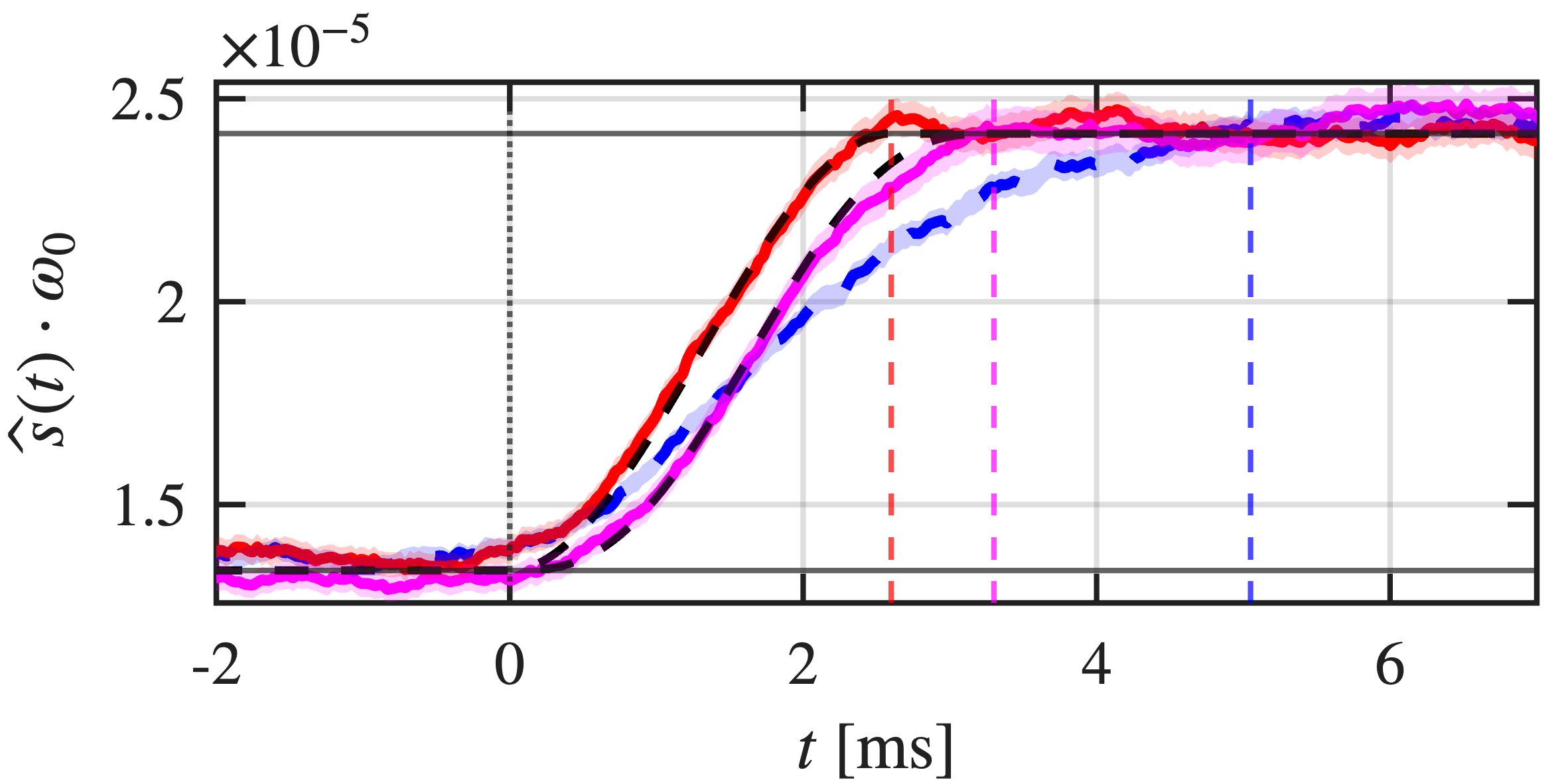}
    \includegraphics[width=0.45\textwidth]{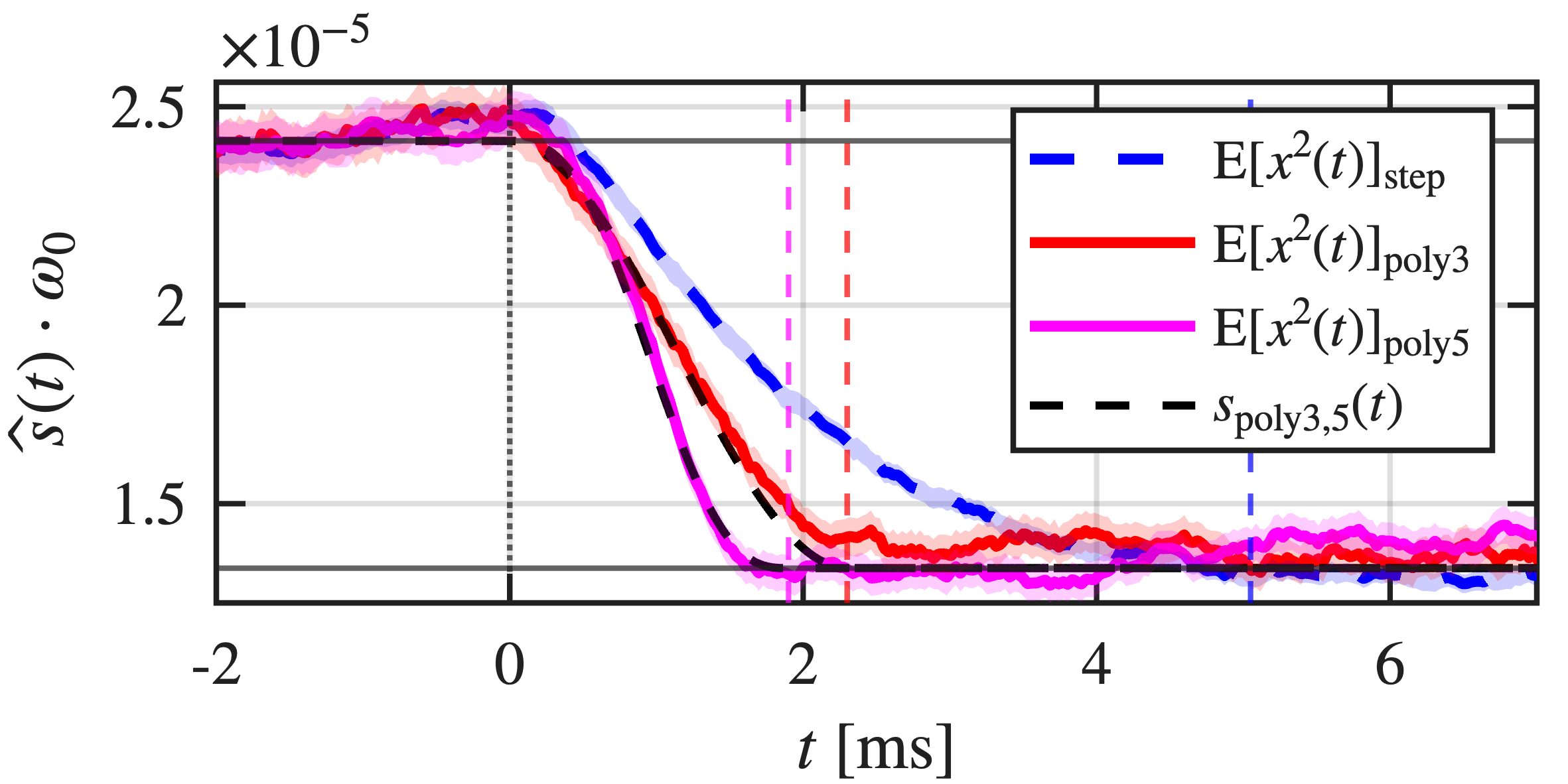}
    \captionsetup{justification=raggedright,singlelinecheck=false}
    \caption{
    Accelerated NESS transitions.
    Noise-color protocols $\omega_c(t)$ (upper row, solid curves) and corresponding normalized ensemble-averaged variances $\hat{s}(t)$ obtained from experimental trajectories (lower row), compared with an instantaneous STEP protocol (blue dashed curves).
    Black dashed curves show the analytical predictions of Eqs.~\eqref{eq:poly3} and \eqref{eq:poly5}, and shaded regions indicate standard errors.    
    Horizontal lines indicate the initial and final NESS values $\hat{s}_{i,f}$ (Eq.~\eqref{eq:NESS}).
    \textbf{(a)} Expansions $\Delta\hat{s}>0$, employing third and fifth-order protocols ($p_3$, $p_5^{\mathrm{cts}}$), with $t_{f,1}=2.60~\mathrm{ms}$ and $t_{f,2}=3.30~\mathrm{ms}$ respectively. 
    \textbf{(b)} Compressions $\Delta\hat{s}<0$, with $t_{f,1}=2.30~\mathrm{ms}$ and $t_{f,2}=1.90~\mathrm{ms}$.
    Both compared to a STEP relaxation ($\tau_{\mathrm{step}}\approx5~\mathrm{ms}$). 
    Experimental parameters are as in Fig.~\ref{fig:fig2}.
    }
    \label{fig:main_fig}
\end{figure*}

The boundary conditions \eqref{eq:sif}, together with the positivity constraints on $z(t)$ and $\omega_c(t)$, completely characterize the admissible variance trajectories. 
For a fixed trajectory shape, they impose a speed limit (i.e., a minimal admissible transition duration)
\begin{equation}
\begin{split}
t_f^\mathrm{min} \defeq \inf \{\, t_f > 0 :\;
z(t) > 0, \; \omega_c(t) > 0 \;\forall t \in [0,t_f] \,\}
\end{split},
\label{eq:tf_min}
\end{equation}
and the infimum of $t_f^\mathrm{min}$ over all trajectory shapes establishes the fundamental speed limit for finite-time steering by noise color. 
The speed limit $t_f^\mathrm{min}$ must generally be computed numerically; however, for strictly monotone trajectories, Sec.~II of the SM \cite{SM} gives analytical lower and upper bounds.

Next, we derive explicit formulas for $\omega_c(t)$ by prescribing polynomial forms for the variance $s(t)$, which satisfy the boundary conditions \eqref{eq:sif}. 
The resulting protocols are fully analytical, experimentally implementable, and allow for precise control of the transition duration $t_f$. 

{\it Third-order polynomial}.---The lowest-degree polynomial satisfying the boundary conditions in Eq.~\eqref{eq:sif}  
is 
\begin{equation}
\label{eq:poly3}
s(t)
=
s_i
+
\Delta s \;
\tau^2
\left(
3-2\tau
\right),
\end{equation}
where we have defined $\tau = t/t_f$ and $\Delta s \defeq s_f-s_i$ (cf. Ref.~\cite{Martinez2016}).
Substituting this variance trajectory into the moment equations yields explicit expressions for $z(t)$ and the control protocol $\omega_c(t)$, which we refer to as the third-order protocol $p_3$. The resulting control $\omega_c(t)$ is discontinuous at the boundaries $t=0$ and $t=t_f$, producing instantaneous jumps in the correlation rate (see Fig.~\ref{fig:fig2}(a)).

{\it Class of fifth-order polynomials}.---
By considering a fifth-order polynomial for $s(t)$, we introduce two additional degrees of freedom, yielding a two-parameter class parametrized by $(a_1,a_2)\in\mathbb{R}^2$:
\begin{equation}
\label{eq:poly5}
s(t)
=
s_i
+
\Delta s \;
\tau^2\left[ 3-2\tau
+(1-\tau)^2\left(a_1+a_2(2\tau-1)\right) \right].
\end{equation}
The choice $(a_1,a_2)=(0,0)$ recovers the third-order protocol $p_3$.
The two additional parameters provide greater flexibility in shaping the evolution of $s(t)$ while preserving the same endpoints and transition time.
As shown below, this freedom can be exploited to suppress discontinuities in the control, reduce the speed limit, or optimize a thermodynamic cost.

More generally, the physical admissibility conditions impose affine linear constraints on $(a_1,a_2)$, defining a convex region of feasible parameters (see Sec.~IV of the SM \cite{SM}).
We first consider the smooth protocol $p_5^\mathrm{cts}$, with $(a_1,a_2)=(0,3)$, for which $\omega_c(t)$ is continuous (see Fig.~\ref{fig:main_fig}(a-b)).
Although discontinuities in the control parameter are fully realizable in our overdamped experimental regime, smooth protocols are relevant when such discontinuities are not operational \cite{Chupeau2018}.

\begin{figure}[h!]
    \centering
    \includegraphics[width=0.45\textwidth]{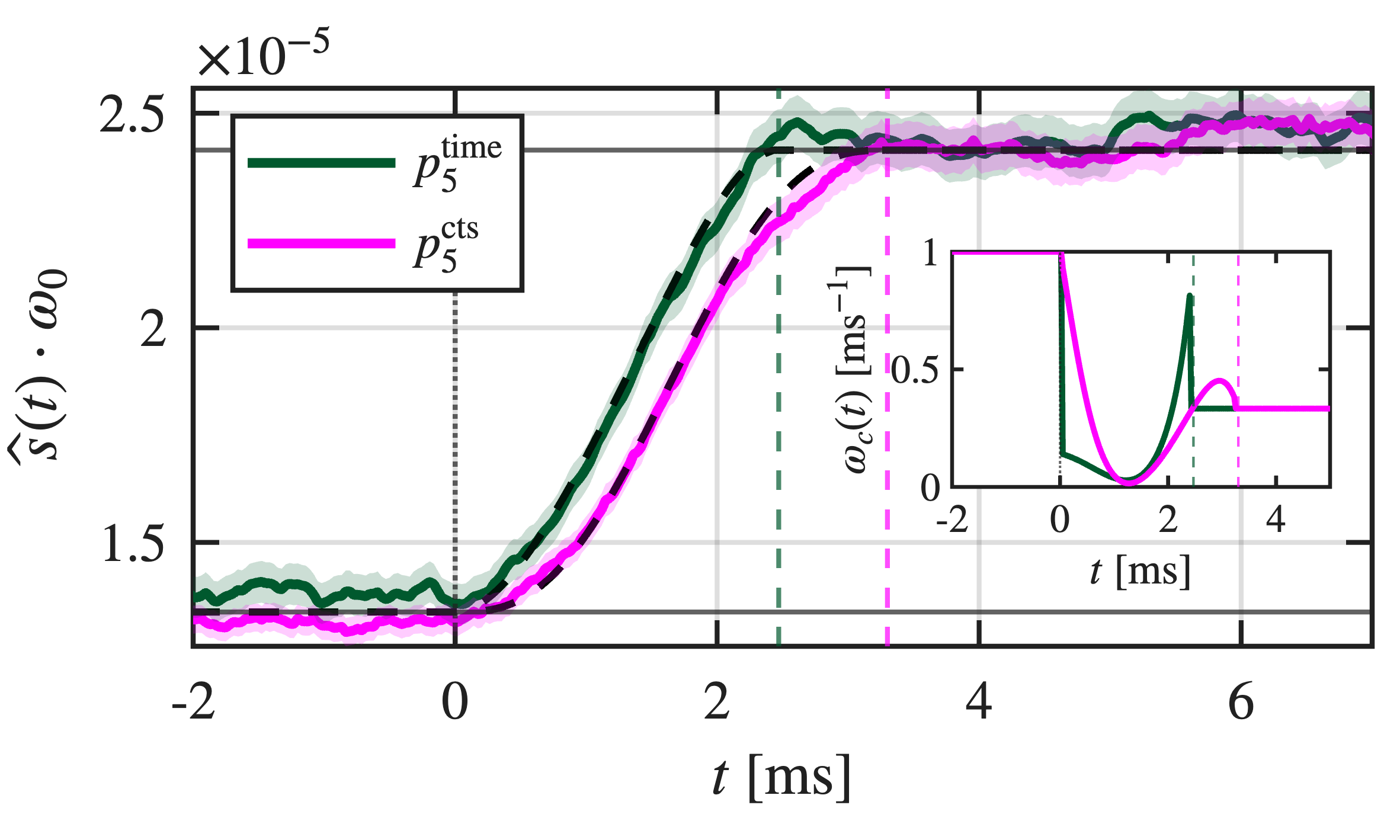}
    \captionsetup{justification=raggedright,singlelinecheck=false}
    \caption{
    Time-optimized fifth-order protocol $p_5^{\mathrm{time}}$ (green), with $(a_1,a_2)=(-0.55,-0.85)$, compared with the continuous protocol $p_5^{\mathrm{cts}}$ (magenta) for an expansion process (cf. Fig.~\ref{fig:main_fig}(a)). 
    Black dashed curves show the analytical prediction of Eq.~\eqref{eq:poly5}, and shaded regions indicate standard errors.    
    The optimized transition time $t_f=2.47$ ms, compared with $t_f=3.30$~ms for $p_5^{\mathrm{cts}}$ (horizontal dashed lines). 
    The inset shows the corresponding control functions $\omega_c(t)$. 
    Both protocols are performed at their speed limit $t_f \gtrsim t_f^{\textrm{min}}$. 
    Experimental parameters are as in Fig.~\ref{fig:fig2}.
    }
    \label{fig:fig3}
\end{figure}

{\it Normalization}.---The active diffusivity $D_a$ is not fully reproducible between experiments (see Sec.~VI.B of the SM \cite{SM}), which makes the results of different experimental trials difficult to compare. This limitation may be circumvented by defining the normalized moments $\hat s=\left( s-D/\omega_0 \right)/(2D_a)$ and $\hat z= z/\sqrt{2D_a}$. These normalized moments obey evolution equations that do not depend on $D_a$, and are thus comparable across experiments (note that $\hat{s}$ has units of time).
The inversion formula \eqref{eq:omegac} is also independent of $D_a$, so a single protocol steers every realization correctly irrespective of differences in $D_a$.

{\it Experimental results}.---
We experimentally demonstrate our control protocol via the active bath engineering of a colloidal particle confined in a harmonic potential.
A $3\mu m$ polystyrene bead immersed in water is optically trapped by a tightly focused laser beam (optical tweezer) and stochastically driven along the beam axis by the radiation pressure exerted by a second, \textit{'pushing'} beam.
The intensity of the latter is modulated with an acousto-optic modulator (AOM) according to an OU process (Eq.~\eqref{eq:eta_process}), thus realizing experimentally the AOUP dynamics of Eq.~\eqref{eq:x_process}.
Here, the control parameter is the active noise color $\omega_c(t)$, modulated by the pushing beam, while the trapping beam, and hence the harmonic relaxation rate $\omega_0$, is kept constant.
Further details of the experimental setup are provided in the Sec.~VI.A of the SM \cite{SM} (cf. Ref.~\cite{Goerlich2022b}).

Fig.~\ref{fig:fig2} shows the experimental implementation of the third-order protocol $p_3$ between NESSs corresponding to $\omega_c=1$~ms$^{-1}$ and $\frac{1}{3}$~ms$^{-1}$, 
with different prescribed transition times $t_f$.
Here, solid curves show normalized ensemble averages of the experimental trajectories, while black dashed curves denote the analytical predictions of Eq.~\eqref{eq:poly3}.
Vertical dashed lines mark the prescribed transition times $t_f$, and horizontal lines indicate the stationary values ($\hat{s}_{i,f}$) from Eq.~\eqref{eq:NESS}.

\begin{figure}[t]
    \centering
    \includegraphics[width=0.45\textwidth]{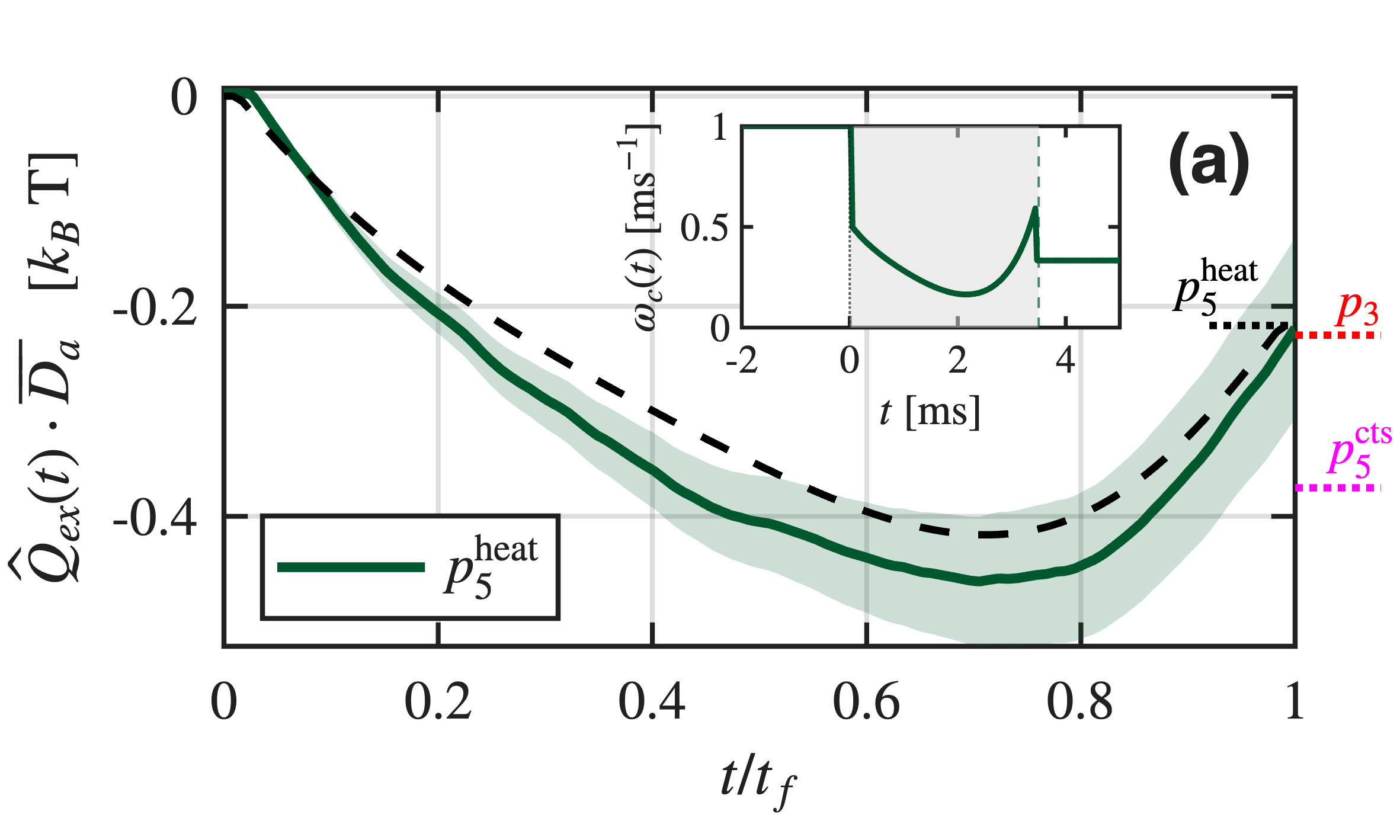}
    \includegraphics[width=0.45\textwidth]{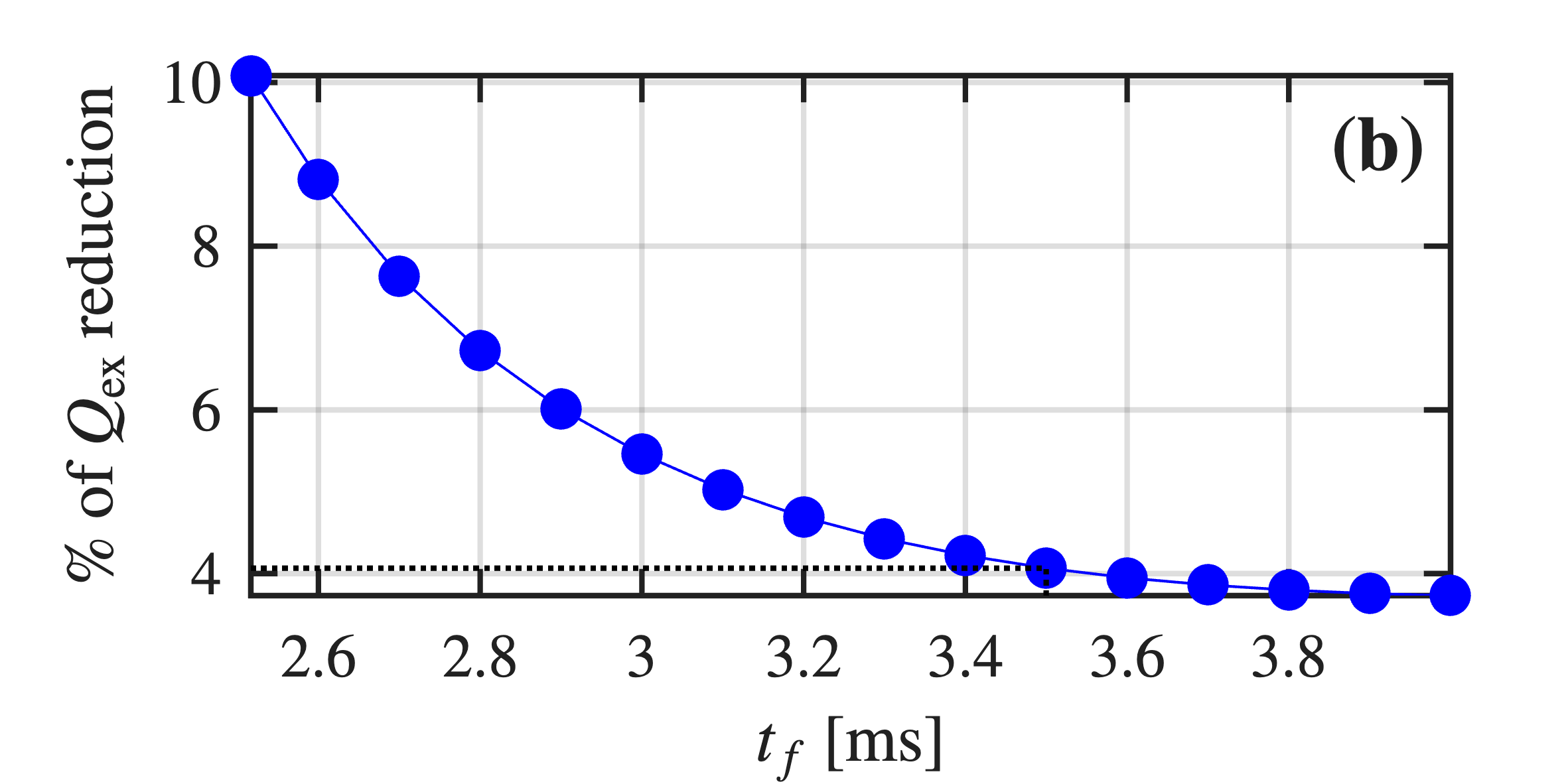}
    \captionsetup{justification=raggedright,singlelinecheck=false}
    \caption{
    \textbf{(a)} 
    Normalized excess heat (Eq.~\ref{eq:Qex_finite_time}) for the heat-optimal protocol $p_5^{\mathrm{heat}}$ with $t_f=3.50$~ms and $(a_1,a_2)=(-0.25,-1.05)$;
    $\overline{D_a}$ is the experimental average active diffusion coefficient (to restore units of energy);
    the inset shows the corresponding control function $\omega_c(t)$. 
    Solid line shows the experimental average, shaded region denotes standard errors, and black dashed line indicates the analytical solution. 
    Horizontal dotted lines compare analytical accumulated excess heats (at $t=t_f$) with $p_3$ and $p_5^{\mathrm{cts}}$.
    Experimental parameters are as in Fig.~\ref{fig:fig2}.
    \textbf{(b)} Excess heat reduction of heat-optimized $p_5^{\mathrm{heat}}$ versus $p_3$ protocols from Monte-Carlo simulations ($M=10^7$), estimated over a range of prescribed transition times $t_f$.
    }
    \label{fig:fig4}
\end{figure}

The protocol provides direct control over the transition duration and, for short $t_f\approx t_f^{\rm min}$, significantly accelerates the NESS-to-NESS transition relative to the natural relaxation following a step change in $\omega_c$; see Sec.~V of SM \cite{SM} for further details of step protocol.
Experimentally, we choose $t_f$ slightly above this limit to avoid very small values of $\omega_c$, since the active OU driving $\eta(t)$ is sensitive to intrinsic AOM noise.

Fig.~\ref{fig:main_fig} focuses on accelerated ($t_f\approx t_f^{\textrm{min}}$) expansions ($\Delta s>0$) and compressions ($\Delta s<0$), comparing the $p_3$ and $p_5^\mathrm{cts}$ protocols with the corresponding step relaxation.
For both processes, the controlled system reaches the target NESS faster than the natural relaxation process.
Notably, for compression $p_5^\mathrm{cts}$ can be accelerated beyond $p_3$ owing to its lower $t_f^{\rm min}$, while for expansion the opposite holds.

{\it Speed limit and thermodynamic cost}.---
Rather than imposing continuity on the fifth-order protocol, its additional degrees of freedom can be exploited to minimize the speed limit.
The infimum of $t_f$ for which the feasible parameter region remains nonempty defines the time-optimized protocol $p_5^\mathrm{time}$ and its speed limit $t_f^\mathrm{min}$ (see Sec.~IV of the SM \cite{SM}).

Fig.~\ref{fig:fig3} demonstrates this optimization experimentally, implementing $p_5^\mathrm{time}$ with optimal $(a_1,a_2)$ under the same experimental conditions as Fig.~\ref{fig:main_fig}.
The time-optimized protocol reaches the target NESS in $t_f=2.47$ ms, compared with $3.3$ ms for $p_5^\mathrm{cts}$, and reduces the speed limit by $5\%$ relative to the corresponding $p_3$ protocol.

The choice of variance trajectory also controls the finite-time thermodynamic response. 
Working with the normalized quantities defined above, we introduce the covariance lag $\hat\delta(t)=\hat z_{\rm ss}(\omega_c(t))-\hat z(t)$, with $\hat z_{\rm ss}=\alpha/(\omega_0+\omega_c)$, which measures how far the covariance lags behind the instantaneous stationary value it would take under an infinitely slow drive. 
In terms of its accumulated value $\Delta(t)=\int_0^t\hat\delta(s)\,ds$, the excess heat reads $\hat Q_{\rm ex}(t_f)=\tfrac{\kappa}{2}(\hat s_f-\hat s_i)-\kappa\Delta(t_f)$, and the second moment dynamics yield the exact decomposition \cite{Oono1998,Hatano2001,Goerlich2022b}
\begin{equation}
\label{eq:Qex_finite_time}
\hat Q_{\rm ex}(t_f)
=
\hat Q_{\rm ex}^{\rm qs}
-\frac{\kappa}{\alpha}
\int_0^{t_f}(\omega_0+\omega_c(s))\hat\delta(s)^2\,ds,
\end{equation}
where (minus) the second term is non-negative and vanishes as $t_f\to\infty$, and $\hat Q_{\rm ex}^{\rm qs}=\frac{\kappa}{2}(\hat s_f-\hat s_i)-\frac{\kappa\omega_0^2}{2\alpha}(\hat s_f^2-\hat s_i^2)$ is the quasistatic excess heat. Note that heat released to the thermal bath is denoted by negative sign following \cite{Sekimoto1998,Goerlich2022b}. See Sec.~III of the SM \cite{SM} for more details on the derivation.

The quasistatic contribution depends only on the endpoints and is therefore common to all admissible protocols connecting the same NESSs.
Within the class \eqref{eq:poly5}, the dependence of $\hat Q_{\rm ex}(t_f)$ on $(a_1,a_2)$ at fixed $t_f$ thus arises entirely from the finite-time lag term. 
Minimizing (minus) this contribution over the feasible parameter region defines the heat-optimized protocol $p_5^{\mathrm{heat}}$.
Notably, this optimization is restricted to the chosen polynomial family and therefore does not constitute a globally optimal control solution \cite{Davis2024,Olsen2025}.

Fig.~\ref{fig:fig4}(a) shows the experimental evaluation of $\hat Q_{\rm ex}(t)$ for $p_5^{\mathrm{heat}}$, which shows a reduction in the accumulated excess heat relative to $p_5^{\mathrm{cts}}$ (dotted line) under the same $t_f$.
The black dashed curve is the analytical solution from Eq.~\eqref{eq:Qex_finite_time}.
For our experimental parameters, however, differences in excess heat among the protocols (e.g. between $p_3-p_5^{\mathrm{heat}}$) remain below the experimental uncertainty and cannot be reliably resolved.

Monte Carlo simulations of the experiment with increased statistical averaging recover the predicted reduction and show that the benefit of heat optimization becomes more pronounced for shorter transition times $t_f$ (see Fig~\ref{fig:fig4}(b)).
Importantly, this optimization requires only a different choice of $p_5$ parameters, without modifying the experimental implementation itself.

{\it Conclusion}.---
We have introduced a framework for the finite-time control of active matter based solely on the temporal correlations of the active fluctuations. 
For an AOUP confined in a harmonic trap, we showed that prescribing the evolution of the variance uniquely determines the required time-dependent correlation rate of the noise through an explicit inversion formula. 
As a result, steering between arbitrary NESSs reduces to the construction of an admissible variance trajectory satisfying simple positivity constraints. 

We propose explicit polynomial protocols that realize such transitions and experimentally implement them using an optically trapped colloidal particle. 
The experiments are in excellent agreement with the theoretical predictions and demonstrate that transitions can be either accelerated or slowed down relative to the standard STEP protocol while keeping the trap stiffness, temperature, and noise amplitude fixed. 
Furthermore, the two additional degrees of freedom of the fifth-order protocol can be exploited to enforce continuity of the control; optimize the speed limit; or minimize a thermodynamic cost. This flexibility naturally motivates the search for a time-energy trade-off relating the achievable protocol duration to its thermodynamic cost.

More broadly, our results establish the correlation time of active fluctuations as a genuine control variable for nonequilibrium systems. 
Although our experimental implementation involves an AOUP, the derivation itself relies only on the closure of the second-moment dynamics (i.e. position variance and position-noise covariance) and is therefore not restricted to the specific standard OU process. 
Beyond the particular AOUP model considered here, the reverse-engineering approach applies to a broad class of Gaussian colored-noise systems and suggests a new route toward the engineering of NESSs through noise design.

\bibliography{bibliography}


\clearpage
\onecolumngrid

\begin{center}
{\Large\bfseries Supplemental Material}
\end{center}

\vspace{0.5em}

\setcounter{secnumdepth}{2}
\setcounter{section}{0}

\section{Further Details on Admissibility of Prescribed Trajectories}
\label{SM:reverse_engineering}

In this section, we provide the remaining technical details on how the conditions imposed on the prescribed trajectory $s(t)$ in the main text are sufficient to ensure its admissibility.

Throughout the Supplemental Materials, we will exploit the equivalence of the protocol $\omega_c(t)$ obtained from the standard and $D,D_a$-normalized moments; a consequence of the equivalence is that the admissibility conditions, heat formula, and optimization targets considered are also equivalent up to normalization. Recall the normalized moments $\hat s=(s-D/\omega_0)/(2D_a)$ and $\hat z=z/\sqrt{2D_a}$, their moment equations
\begin{equation}
\dot{\hat s}=-2\omega_0\hat s+2\hat z,
\qquad
\dot{\hat z}=-\bigl(\omega_0+\omega_c(t)\bigr)\hat z+\alpha.
\label{eq:S_moments}
\end{equation}
and their normalized inversion formulas
\begin{equation}
\hat z=\omega_0\hat s+\tfrac12\dot{\hat s},
\qquad
\omega_c=\frac{\alpha-\dot{\hat z}}{\hat z}-\omega_0.
\label{eq:S_inversion}
\end{equation}
Furthermore, recall the normalized NESS values for fixed $\omega_c$ are
\begin{equation}
\hat s_{\rm ss}(\omega_c)=\frac{\alpha}{\omega_0(\omega_0+\omega_c)}, \qquad \hat z_{\rm ss}(\omega_c)=\frac{\alpha}{\omega_0+\omega_c},
\label{eq:S_NESS}
\end{equation}
and we write $\hat s_{\ii,\ff}=\hat s_{\rm ss}(\omega_{c}(t_{\ii, \ff}))$ and $\hat z_{\ii,\ff}=\hat z_{\rm ss}(\omega_{c}(t_{\ii,\ff}))$ ($t_\ii = 0$).

A prescribed trajectory $\hat s(t)$ is admissible when the reconstructed protocol $\omega_c(t)$ (R1) is physically realizable, (R2) achieves the transition, and (R3) is continuous on $(0,t_\ff)$, with jumps permitted on the boundaries $t=0$ and $t=t_\ff$. We show that it suffices to impose the following conditions on $\hat s(t)$: (C1) $\hat s(t)$ twice-differentiable on $[0,t_\ff]$ with $\ddot{\hat s}(t)$ continuous on $(0,t_\ff)$; (C2) $\hat s(0) = \hat s_\ii$, $\hat s(t_\ff) = \hat s_\ff$, and $\dot{\hat s}(0) = \dot{\hat s}(t_\ff) = 0$; and (C3) $\hat z(t)$ and $\omega_c(t)$ strictly positive. It is clear that by Eq.~\eqref{eq:S_inversion} C1 and C3 imply R3 directly.

For R1, the correlation rate $\omega_c(t)$ must be positive, which follows from C3, and the reconstructed moments must remain a valid Gaussian, i.e., the covariance matrix $ \hat \Sigma(t)$ must remain positive definite for all $t$. Since the determinant obeys the evolution equation
\begin{equation*}
\frac{d}{dt}\det \hat \Sigma(t)
=
-2\omega_0\det \hat \Sigma(t)
+2\omega_c(t) \hat z(t)^2,
\end{equation*}
the positivity of $2\omega_c(t) \hat z(t)^2$ and the initial condition $\hat \Sigma(0) > 0$, which follow from C3 and a valid initial NESS, result in $\det \hat \Sigma(t) \geq \det \hat \Sigma(0) e^{-2\omega_0t} > 0$.

For R2, the reconstructed moments must take their stationary values and be stationary at the endpoints. For $\hat s$, this follows immediately from C2, while for $\hat z$, we also have $\hat z(0)=\hat z_\ii$ and $\hat z(t_\ff)=\hat z_\ff$ by Eq.\eqref{eq:S_inversion}. Since we allow $\omega_c(t)$ to jump at the endpoints, $\hat z$ need not be differentiable at the endpoints. Instead, we show that the one-sided derivatives vanish, $\dot{\hat z}(0^-) = \dot{\hat z}(t_\ff^+) = 0$, which is sufficient for our purposes. Indeed, since $\omega_c$ remains at the values $\omega_{c,\ii}$, $\omega_{c,\ff}$ before and after the protocol, we have that Eq.~\eqref{eq:S_moments} gives $\dot{\hat z}(0^-)=-(\omega_0+\omega_{c,\ii})\hat z_\ii+\alpha=0$ and $\dot{\hat z}(t_\ff^+)=0$.

\section{Minimal Admissible Duration (Speed Limit) and Closed-Form Bounds}

In the main text, each prescribed trajectory $\hat s(t)$ satisfies C1 and C2 by construction, so its admissibility is solely determined by C3, which depends on the transition duration $t_\ff$. This $t_\ff$-dependence gives rise to the minimal admissible duration (or speed limit) $t_\ff^\mathrm{min}$ as detailed in the main text. In general, no analytical expression exists for $t_\ff^\mathrm{min}$, and it must be computed numerically. Here, we derive closed-form upper and lower bounds, which correspond to sufficient and necessary conditions for C3, respectively. We first identify $\hat s(t)$ with its unit shape function $p(\tau)$, $\tau \in [0,1]$, which satisfies $p(0) = 0$, $p(1) = 1$, and $p'(0)=p'(1)=0$:
\begin{equation}
\hat s(t)=\hat s_\ii+\Delta\hat s\,p(t/t_\ff),
\qquad
\Delta\hat s=\hat s_\ff-\hat s_\ii.
\label{eq:S_decomposition}
\end{equation}
For the third and fifth-order protocols in the main text, these are $p_3(\tau)=3\tau^2-2\tau^3$, $p_5^\mathrm{cts}(\tau)=\tau^3 (10-15\tau+6\tau^2)$, and $p_5(\tau;a_1,a_2)$ for general $(a_1,a_2)$. Suppose that $p$ is strictly increasing, as is the case for $p_3$, $p_5^\mathrm{cts}$, and $p_5^\mathrm{time}$, and denote the derivatives with respect to $\tau$ by primes. Substituting Eq.~\eqref{eq:S_inversion} into C3, we obtain after rearrangement the forms
\begin{align}
\hat z&=\hat z_\ii+\Delta\hat s\Bigl[\omega_0p+\frac{p'}{2t_\ff}\Bigr]
\label{eq:S_shape_z}\\
\omega_c\hat z&=\omega_{c,\ii}\hat z_\ii-\Delta\hat s\Bigl[\omega_0^2p+\frac{3\omega_0p'}{2t_\ff}+\frac{p''}{2t_\ff^2}\Bigr]
\label{eq:S_shape_wz}
\end{align}
whose positivities $\hat z > 0$ and $\omega_c \hat z > 0$ on $t \in [0,t_\ff]$ are equivalent to that of C3, whose signs are independent of $\alpha$, and which is affine in $p$. The minimal admissible duration $t_\ff^\mathrm{min}$ is hence only dependent on the shape $p$ and parameters $\omega_0$, $\omega_{c,\ii}$, and $\omega_{c,\ff}$.

\paragraph{Lower bound on $t_\ff^\mathrm{min}$.} A necessary condition for the positivity of $\hat z$ and $\omega_c \hat z$ is positivity only at the interior limiting endpoints $\tau \rightarrow 0^+$ and $\tau \rightarrow 1^-$. After algebraic manipulations, this gives the lower bound:
\begin{equation}
(t_\ff^\mathrm{min})^2 \geq \max\Bigl\{\frac{\Delta\hat s\,p''(0)}{2\omega_{c,\ii}\hat z_\ii},\ \frac{\Delta\hat s\,p''(1)}{2\omega_{c,\ff}\hat z_\ff},\ 0\Bigr\}.
\label{eq:S_necessary}
\end{equation}

\paragraph{Upper bound on $t_\ff^\mathrm{min}$.} A sufficient condition for $\hat z > 0$ and $\omega_c \hat z > 0$ on $t \in [0,t_\ff]$ is obtained by replacing the time dependent $p, p', p''$ in Eq.~\eqref{eq:S_shape_z}-\eqref{eq:S_shape_wz} by their extremal values over $\tau \in [0,1]$; whether the maximum or minimum is taken is determined by the sign of $\Delta \hat s$, which differs between an expansion and a compression. In a compression $\Delta \hat s < 0$, after algebraic manipulations, this gives the bound
\begin{equation}
t_\ff^{\rm min}\le\max\left\{\frac{|\Delta\hat s|\max p'}{2\hat z_\ff},\ \left(\frac{|\Delta\hat s|\,|\min p''|}{2\omega_{c,\ii}\hat z_\ii}\right)^{1/2}\right\}.
\label{eq:S_sufficient_comp}
\end{equation}
In an expansion $\Delta \hat s > 0$, since $\hat z\ge\hat z_\ii > 0$, only $\omega_c\hat z>0$ constrains the transition duration. After replacing with the extremal values, we obtain a quadratic inequality in $1/t_\ff$, whose positive root then gives the upper bound:
\begin{equation*}
t_\ff^{\rm min}\le \left(\frac{\bigl[9\omega_0^2(\max p')^2+8\max p''\,\omega_{c,\ff}\hat z_\ff/\Delta\hat s\bigr]^{1/2}-3\omega_0\max p'}{2\max p''} \right)^{-1}.
\end{equation*}
Separately, for an expansion, we can furthermore use an AM-GM inequality on $\omega_0^2p=\omega_0^2-\omega_0^2(1-p)$ in Eq.~\eqref{eq:S_shape_wz} to obtain the alternative upper bound:
\begin{equation}
(t_\ff^{\rm min})^2\le \frac{\Delta\hat s}{2\omega_{c,\ff}\hat z_\ff}\,\sup_{0\le\tau<1}\Bigl[p''+\frac{9\,p'^2}{8(1-p)}\Bigr],
\label{eq:S_sufficient_exp}
\end{equation}
which is tighter than the quadratic bound unless $\Delta \hat s$ is small. The minimal admissible time is hence bounded by the smaller of the two. Note that all bounds can be written explicitly in terms of correlation rates $\omega_{c,\ii}$ and $\omega_{c,\ff}$ by Eq.~\eqref{eq:S_NESS}.
 
Table~\ref{tab:S_speed} compares these bounds of $t_\ff^{\rm min}$ using the parameters from the experiments. The lower bound~\eqref{eq:S_necessary} is attained for $p_3$ in compression and for $p_5^{\rm time}$ in expansion, while the bounds in other cases remain relatively close.
 
\begin{table}[h!]
    \centering
    \renewcommand{\arraystretch}{1.3}
    \setlength{\tabcolsep}{10pt}
    \begin{tabular}{l c c c c c c}
        \hline
        & \multicolumn{3}{c}{Expansion ($\Delta \hat s > 0$)} & \multicolumn{3}{c}{Compression ($\Delta \hat s < 0$)} \\
        \cline{2-7}
        Protocol & Lower \eqref{eq:S_necessary} & $t_\ff^\mathrm{min}$ & Upper \eqref{eq:S_sufficient_exp} & Lower \eqref{eq:S_necessary} & $t_\ff^\mathrm{min}$ & Upper \eqref{eq:S_sufficient_comp} \\
        \hline
        $p_3$ & $2.227$ & $2.533$ & $3.205$ & $2.227$ & $2.227$ & $2.867$ \\
        $p_5^\mathrm{cts}$ & $0$ & $3.251$ & $3.292$ & $0$ & $1.836$ & $2.812$ \\
        $p_5^\mathrm{time}$ & $2.332$ & $2.332$ & $3.885$ & & & \\
        Step & \multicolumn{3}{c}{$6.073$} & \multicolumn{3}{c}{$5.102$} \\
        \hline
    \end{tabular}
    \caption{Bounds on the minimal admissible duration $t_\ff^\mathrm{min}$ (ms), for $\omega_0=0.49~\mathrm{ms}^{-1}$, with $\omega_c:1\to1/3~\mathrm{ms}^{-1}$ (expansion) and $1/3\to1~\mathrm{ms}^{-1}$ (compression); bounds are rounded to 3 d.p. The step row gives the step transition duration of Sec.V. Note that $p_5^\mathrm{time}$ is optimized specifically for an expansion, so compression results are omitted.}
    \label{tab:S_speed}
\end{table}

\section{Excess Heat Derivation}

In this section, we derive the total heat and the non-adiabatic housekeeping-excess decomposition during active noise protocol-induced transitions between NESS. Following Sekimoto \cite{Sekimoto1998, Sekimoto2010}, the heat exchanged with the thermal bath is the work done by the bath forces. By the force balance from the dynamics of $x$, we have along a trajectory the total heat
\begin{equation*}
dq_{\rm tot}=\kappa\,x\circ dx-\gamma\sqrt{2D_a}\,\eta\circ dx,
\end{equation*}
where $\circ$ denotes the Stratonovich product and heat transferred to the bath takes negative sign following the convention of \cite{Goerlich2022b}. The first term corresponds to the change in potential energy $U=\tfrac12\kappa x^2$ while the second term denotes minus the work done by the active force. Taking the ensemble average and noting $\langle\eta^2\rangle=\alpha$, we obtain the ensemble-averaged total heat
\begin{equation}
Q_{\rm tot}(t)=\frac{\kappa}{2}\bigl[s(t')\bigr]_0^t+\sqrt{2D_a}\,\kappa\int_0^t z\,dt'-2D_a\gamma\alpha\,t.
\label{eq:S_Qtot}
\end{equation}
The non-adiabatic housekeeping heat, in the sense of \cite{Oono1998, Goerlich2022b}, is the total heat required to maintain a NESS at an instantaneous value $\omega_c$. At fixed $\omega_c$, the moments take their stationary values $s=s_{\rm ss}(\omega_c)$ and $z=z_{\rm ss}(\omega_c)=\sqrt{2D_a}\,\alpha/(\omega_0+\omega_c)$, so Eq.~\eqref{eq:S_Qtot} reduces to an integral of the heat rate $J(\omega_c)=2D_a\gamma\alpha\,[\omega_0/(\omega_0+\omega_c)-1]$. Integrating the heat rate over the protocol yields the housekeeping heat
\begin{equation*}
Q_{\rm hk}(t)=\int_0^t J\bigl(\omega_c(t')\bigr)\,dt'=-2D_a\gamma\alpha\int_0^t\frac{\omega_c(t')}{\omega_0+\omega_c(t')}\,dt',
\end{equation*}
which reduces to Eq.~(G7) of \cite{Goerlich2022b} for a step protocol. The excess heat defined by the difference $Q_{\rm ex} = Q_{\rm tot} - Q_{\rm hk}$ following \cite{Oono1998} is then
\begin{equation*}
Q_{\rm ex}(t)=\frac{\kappa}{2}\bigl[s(t')\bigr]_0^t-\sqrt{2D_a}\,\kappa\int_0^t\delta(t')\,dt',
\qquad
\delta(t)=z_{\rm ss}\bigl(\omega_c(t)\bigr)-z(t),
\end{equation*}
which after dividing by $2D_a$ gives the normalized excess heat of the main text.
In the normalized form, using the identity $\dot{\hat z}=(\omega_0+\omega_c)\hat\delta$ with $\hat \delta = \delta / \sqrt{2D_a}$, we can further split $\int_0^{t}\hat\delta(t')\,dt'$ into the difference of a boundary and non-negative term
\begin{equation*}
\hat Q_{\rm ex}(t)=\frac{\kappa}{2}\left[\hat s(t') - \tfrac{1}{\alpha} {\hat z(t')}^2 \right]_0^t - \frac{\kappa}{\alpha}
\int_0^{t}
(\omega_0+\omega_c(t'))\hat\delta(t')^2\,dt'.
\end{equation*}
In an infinitely slow quasi-static transition, the excess heat reduces to the first term, which we identify as the quasi-static excess, which is independent of the protocol. This motivates the use of the protocol-dependent second term as the simplified objective function for excess heat-minimization:
\begin{equation}
    \Delta \hat Q_{ex}(t_\ff) = \frac{\kappa}{\alpha} \int_0^{t_\ff}
(\omega_0+\omega_c(t))\hat\delta(t)^2\,dt \geq 0.
    \label{eq:heat_objective}
\end{equation}

\section{Optimized fifth-order protocols}

The class of fifth-order polynomial trajectories is given by
\begin{equation}
p_5(\tau;a_1,a_2)=3\tau^2-2\tau^3+\tau^2(1-\tau)^2\bigl[a_1+a_2(2\tau-1)\bigr],
\end{equation}
which satisfies the boundary conditions for every $(a_1,a_2)$. In this form, the third-order polynomial and continuous fifth-order polynomial trajectories are $p_3=p_5(\cdot\,;0,0)$ and $p_5^{\rm cts}=p_5(\cdot\,;0,3)$, respectively. The choice of coefficients $(a_1,a_2)$ is subject to the positivity constraints Eq.\eqref{eq:S_shape_z}--\eqref{eq:S_shape_wz} but is otherwise free. In this section, we utilize this freedom to 1) shorten the minimal admissible duration and 2) minimize the excess heat objective Eq.~\eqref{eq:heat_objective}.

\begin{figure}[h!]
    \centering
    \includegraphics[width=\textwidth]{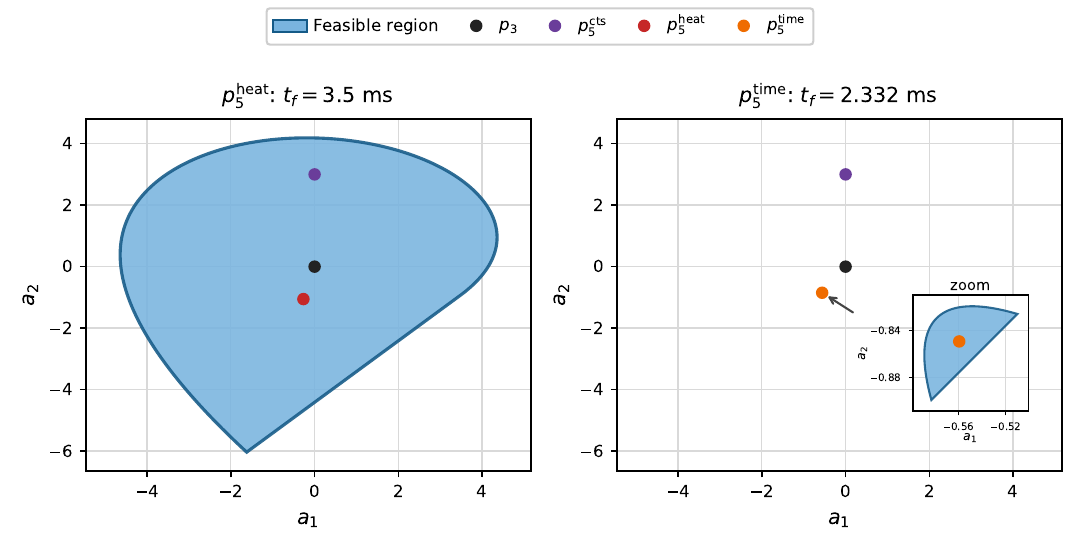}
    \captionsetup{justification=raggedright,singlelinecheck=false}
    \caption{Feasible regions of the fifth-order coefficients for the expansion $\omega_c:1\to1/3~\mathrm{ms}^{-1}$ at $\omega_0=0.49~\mathrm{ms}^{-1}$.
    The feasible region $\mathcal F(t_\ff)$ of Eq.~\eqref{eq:S_feasible} (blue) for the heat optimization at $t_\ff=3.5~\mathrm{ms}$ (left) and the time optimization at $t_\ff=2.332~\mathrm{ms}$ (right). Markers show $p_3$ (black), $p_5^{\rm cts}$ (purple), $p_5^{\rm heat}(3.5)$ (red, left), and $p_5^{\rm time}$ (orange, right). The inset zooms in on the right region around $p_5^{\rm time}$.
    }
    \label{fig:SM_feasible_regions}
\end{figure}

First, for fixed $t_\ff$, we define the feasible region of coefficients $(a_1,a_2)$ that correspond to admissible protocols. In particular, at fixed $t_\ff$, the feasible region $\mathcal{F}(t_\ff)$ is the intersection over the continuum of $(a_1,a_2)$-affine conditions $\hat z(t) > 0$ and $\omega_c(t) \hat z(t)$ for $t \in [0,t_\ff]$. Each condition restricts the coefficients $(a_1,a_2)$ to an open half-plane, and, together, their intersection forms the convex set of admissible coefficients 
\begin{equation}
\mathcal F(t_\ff)=\bigl\{(a_1,a_2):\ \hat z(t)>0\ \text{and}\ \omega_c(t)\hat z(t)>0\ \text{for all}\ t\in[0,t_\ff]\bigr\},
\label{eq:S_feasible}
\end{equation}
Fig.~\ref{fig:SM_feasible_regions} shows the feasible regions for the following excess heat and transition time optimization problems.

\paragraph{Heat optimization.}
At fixed $t_\ff$, finding the heat-optimized protocol $p_5^{\rm heat}(t_\ff)$ is equivalent to the optimization problem of minimizing the excess heat objective function Eq.~\eqref{eq:heat_objective} over the feasible region of $(a_1,a_2)$ at time $t_\ff$:
\begin{equation}
\min_{(a_1,a_2)\in{\mathcal F(t_\ff)}}\ \Delta\hat Q_{\rm ex}(t_\ff;a_1,a_2).
\label{eq:S_heat_problem}
\end{equation}
We solve this numerically. Since $\hat z$ and $\dot{\hat z}$ are affine in $(a_1,a_2)$, the objective is an explicit function of the coefficients, which we evaluate by Gauss--Legendre quadrature. Table~\ref{tab:S_optimized} presents the minimized coefficients $(a_1,a_2)$, accurate to 6 decimal places in the sense that doubling the quadrature nodes does not change the value. For illustration, Fig.~\ref{fig:SM_feasible_regions} shows the numerically approximated feasible region on which we minimize over and the optimized point corresponding to the heat-optimized protocol $p_5^{\rm heat}$.

\paragraph{Time optimization.}
To find the time-optimized protocol $p_5^{\rm time}$, we solve a slightly different minimization problem. Since shortening the transition increases $\dot{\hat s}$ and $\ddot{\hat s}$ in $\hat z$ and $\omega_c\hat z$, the feasible region $\mathcal F(t_\ff)$ becomes empty below some duration. Hence, ideally, the time-optimized protocol $p_5^{\rm time}$ should correspond to a point in the intersection over all non-empty feasible region, giving the shortest attainable duration
\begin{equation}
t_\ff^{\rm min}(p_5)=\inf_{(a_1,a_2)}t_\ff^{\rm min}\bigl(p_5(\cdot\,;a_1,a_2)\bigr)=\inf\bigl\{t_\ff>0:\ \mathcal F(t_\ff)\neq\emptyset\bigr\}.
\label{eq:S_time_problem}
\end{equation}
However, in practice, such a set can be empty and, more importantly, transition durations are limited by machine precision and experimental limitations. Therefore, we instead find the shortest duration $t_\ff$ rounded to 3 d.p. that corresponds to a non-empty feasible region. In particular, we find that $\mathcal F(2.332~\mathrm{ms})$ is non-empty, whereas $\mathcal F(2.331~\mathrm{ms})$ is empty, so $p_5^{\rm time}$ is chosen to be a point in $\mathcal F(2.332~\mathrm{ms})$; its values are reported in Table~\ref{tab:S_speed} and Fig.~\ref{fig:SM_feasible_regions} shows the feasible region.

\begin{table}[h!]
    \centering
    \renewcommand{\arraystretch}{1.3}
    \setlength{\tabcolsep}{8pt}
    \begin{tabular}{l c c c c}
        \hline
        Protocol & $(a_1,a_2)$ & Optimized value & $p_3$ & $p_5^{\rm cts}$ \\
        \hline
        $p_5^{\rm time}$ & $(-0.559444,\,-0.849080)$ & $t_\ff=2.332$~ms & $2.533$~ms & $3.251$~ms \\
        $p_5^{\rm heat}(3.5)$ & $(-0.266959,\,-1.055153)$ & $\Delta\hat Q_{\rm ex}=3.326\times10^{-14}$ & $3.469\times10^{-14}$ & $5.413\times10^{-14}$ \\
        \hline
    \end{tabular}
    \caption{Optimized fifth-order protocols for the expansion $\omega_c:1\to1/3~\mathrm{ms}^{-1}$ at $\omega_0=0.49~\mathrm{ms}^{-1}$. The second column gives the coefficients rounded to six decimals, the third the optimized quantity, and the last two the same quantity for $p_3$ and $p_5^{\rm cts}$: the minimal admissible duration (Table~\ref{tab:S_speed}) for $p_5^{\rm time}$, and $\Delta\hat Q_{\rm ex}$ of Eq.~\eqref{eq:heat_objective} at $t_\ff=3.5$~ms with $\alpha=0.01$ for $p_5^{\rm heat}(3.5)$. $p_5^{\rm heat}(3.5)$ is used in the experiment of Fig.4(a) in the main text.}
    \label{tab:S_optimized}
\end{table}

\section{Step Protocol Statistics and Transition Duration}

In the main text, we compare the reverse-engineered protocols with the step protocol of \cite{Goerlich2022b} as a benchmark. Here, we formally introduce the protocol, the corresponding statistics, and the definition of its transition duration.

The step protocol is an instantaneous change of the correlation frequency at $t=0$, represented by
\begin{equation*}
    \omega_c(t) = \begin{cases} \omega_{c,\ii}, & t \leq 0, \\ \omega_{c,\ff}, & t > t_\ff \end{cases}.
\end{equation*}
Unlike the reverse-engineered finite-time protocols, the system under a step protocol relaxes autonomously toward the target NESS, given by the following analytical solution of its moment
\begin{equation*}
    s(t) = s_\ff + \frac{(s_\ii - s_\ff)}{\omega_0 - \omega_{c,\ff}} \left( 2\omega_0 e^{-(\omega_0+\omega_{c,\ff})t} - (\omega_0+\omega_{c,\ff}) e^{-2\omega_0t} \right).
\end{equation*}
Since the relaxation is governed by two exponential decay rates $2\omega_0$ and $\omega_0+\omega_{c,\ff}$, the system's characteristic time is controlled by the slower rate. This motivates the following definition of the step transition duration:
\begin{equation*}
    \tau_{\mathrm{step}} = 5/\min\left\{2\omega_0, \omega_0+\omega_{c,\ff}\right\}.
\end{equation*}
At $t=\tau_\mathrm{step}$, the slowest exponential has decayed to $e^{-5} \approx 6.7 \times 10^{-3}$. Thus, assuming $\omega_{c,\ff}$ and $\omega_0$ are not close, this transition duration serves as a rough measure of the system's relaxation to the target NESS.

\section{Experimental details}

\subsection{Optical setup}

\begin{figure}[h]
    \centering
    \includegraphics[width=0.52\textwidth]{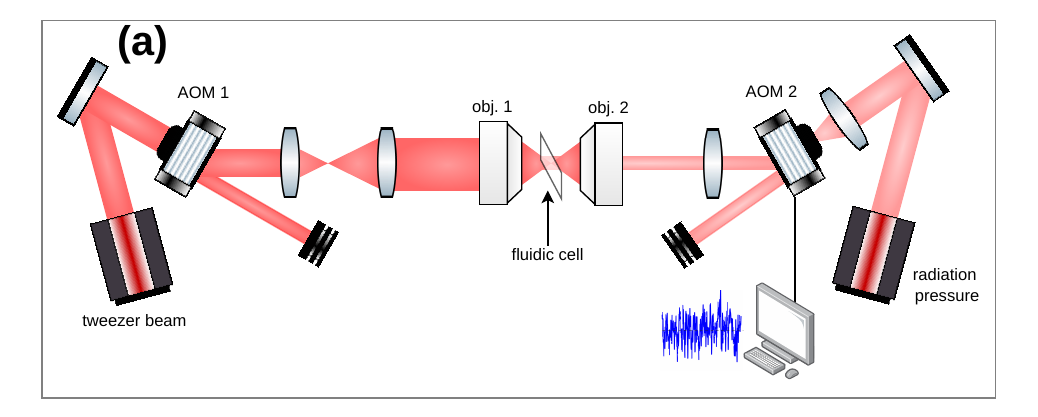}
    
    \includegraphics[width=0.37\textwidth]{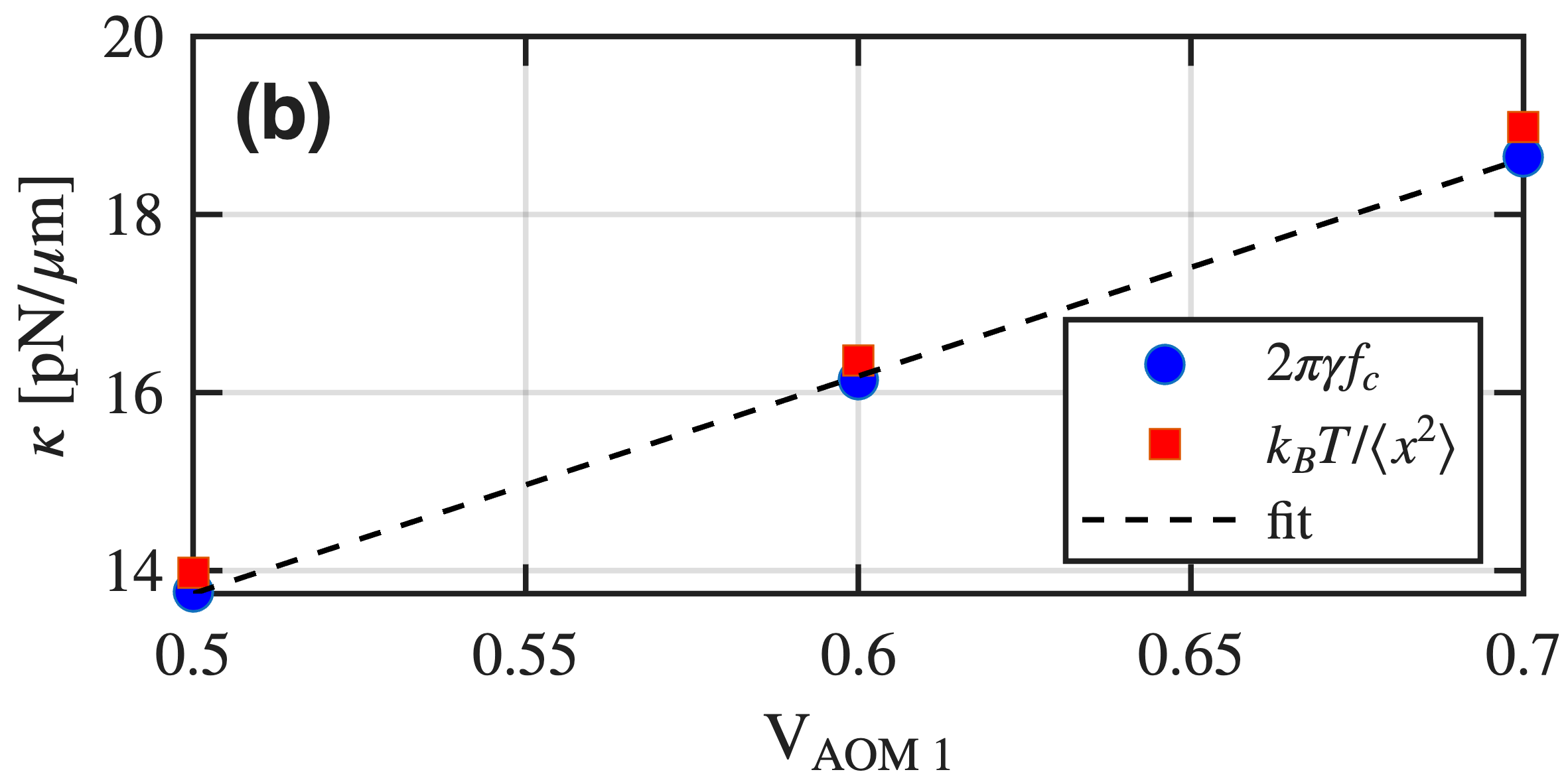}
    \includegraphics[width=0.37\textwidth]{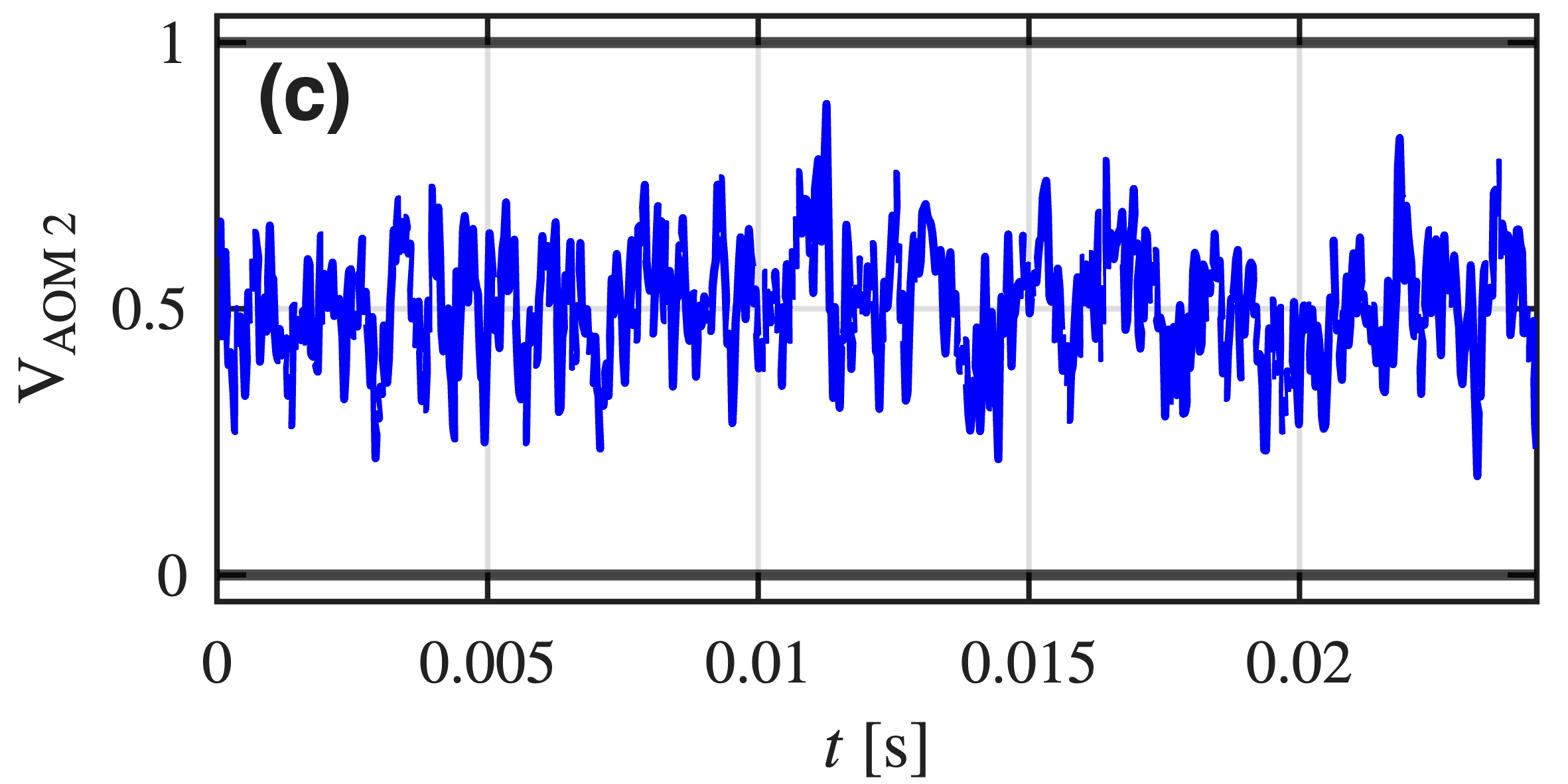}    
    \captionsetup{justification=raggedright,singlelinecheck=false}
    \caption{
    \textbf{Experimental setup:} \ 
    \textbf{(a)} The trapping beam (left), controlled by AOM(1), is focused by the high-NA objective (obj.~1) to confine a colloidal particle inside the fluidic cell. A second pushing beam (right), controlled by AOM(2) and focused by obj.~2, exerts radiation pressure on the particle. AOM(2) is driven by a computer-generated stochastic signal, realizing the active forcing.
    \textbf{(b)} The stiffness $\kappa$ is measured as a function of the trapping-beam control voltage $V_{\mathrm{AOM,1}}$ using two independent calibration methods (symbols):
    corner frequency of the power-spectrum density, $f_c$ (circles); 
    energy equipartition via the mean potential energy, $\frac{1}{2}k_BT=\frac{1}{2}\kappa \langle x^2\rangle$ (squares).
    The dashed line shows the linear fit to the measurements.
    \textbf{(c)} An Ornstein-Uhlenbeck voltage signal $V_{\mathrm{AOM,2}}(t)$ of AOM(2), modulating the radiation-pressure force exerted by the pushing beam and driving the particle into a NESS.
    }
    \label{fig:exp_SM1}
\end{figure}

Our experimental setup, sketched in Fig.~\ref{fig:exp_SM1}\textbf{(a)}, consists of two optical objectives: a high-numerical-aperture (NA) water-immersion objective (1) (Nikon Plan Apochromat, $\times60$, $\mathrm{NA}=1.2$) and a lower-NA air objective (2) (Nikon Plan Fluor, $\times60$, $\mathrm{NA}=0.7$). A Ti:Sapphire laser (Spectra-Physics 3900S), operating at a wavelength of $800~\mathrm{nm}$, is split into two beams: a trapping beam that confines the particle and a pushing beam that exerts radiation pressure on it. For clarity, the two beams are depicted as separate laser sources in Fig.~\ref{fig:exp_SM1}\textbf{(a)}.

The trapping beam is expanded to overfill the back aperture of objective (1), which tightly focuses the beam onto the sample. The sample consists of a $17.5~\mu\mathrm{L}$ liquid suspension of $3~\mu\mathrm{m}$-diameter polystyrene microspheres (Duke Scientific Corp.), confined between a microscope slide and a coverslip separated by a $120~\mu\mathrm{m}$-thick Grace Bio-Labs spacer. The colloidal suspension is diluted in water to a concentration of $0.5\times10^{-4}\%$ (volume fraction).
The room temperature is $T\approx 295$~K, and the viscosity of water is computed as $\tilde{\eta}=0.001$~Pa$\cdot$s so that the Stokes-Einstein coefficient is $D=k_BT/6\pi R\tilde{\eta}$, where the particle radius is $R=1.5$~$\mu$m.

The intensity of the trapping beam, and hence the stiffness $\kappa$ of the harmonic optical potential, is controlled by an acousto-optic modulator (AOM(1)). Figure~\ref{fig:exp_SM1}\textbf{(b)} shows the calibration of $\kappa$ as a function of the driving voltage applied to AOM(1), performed in the absence of active-noise modulation. The measurements are restricted to the linear response regime of the AOM. Throughout the experiments reported in this work, the trap stiffness is held fixed at $\kappa=14~\mathrm{pN}/\mu\mathrm{m}$.

The external stochastic forcing is generated by a second, lower-power divergent beam (the pushing beam), which exerts a radiation-pressure force along its propagation axis. Its intensity is controlled by a second AOM(2), driven through a data-acquisition board (National Instruments PCI-6251) at a sampling rate of $32,768~\mathrm{Hz}$. The driving signal is generated numerically in Python as a discrete stochastic time series with output voltages in the range $[0,1]~\mathrm{V}$. In the experiments presented in the main text, this signal follows an Ornstein-Uhlenbeck process, as illustrated in Fig.~\ref{fig:exp_SM1}\textbf{(c)}, with the correlation rate $\omega_c(t)$ as a time-dependent control parameter, and a constant variance $\alpha\equiv\langle \eta^2\rangle=0.01$~Hz for which the signal does not exceed the AOM range.

\begin{figure}[h]
    \centering
    \includegraphics[width=0.37\textwidth]{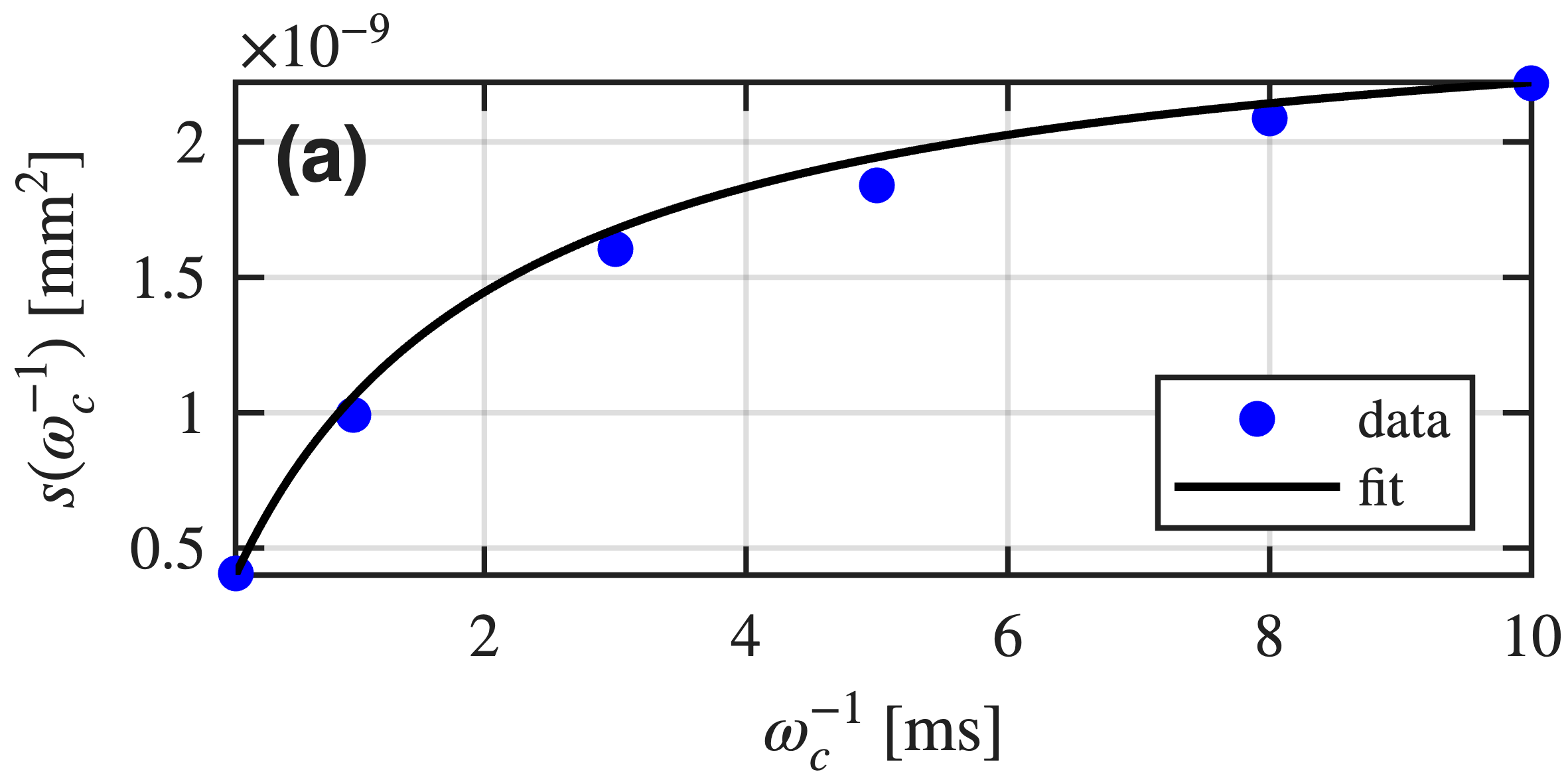}
    \includegraphics[width=0.37\textwidth]{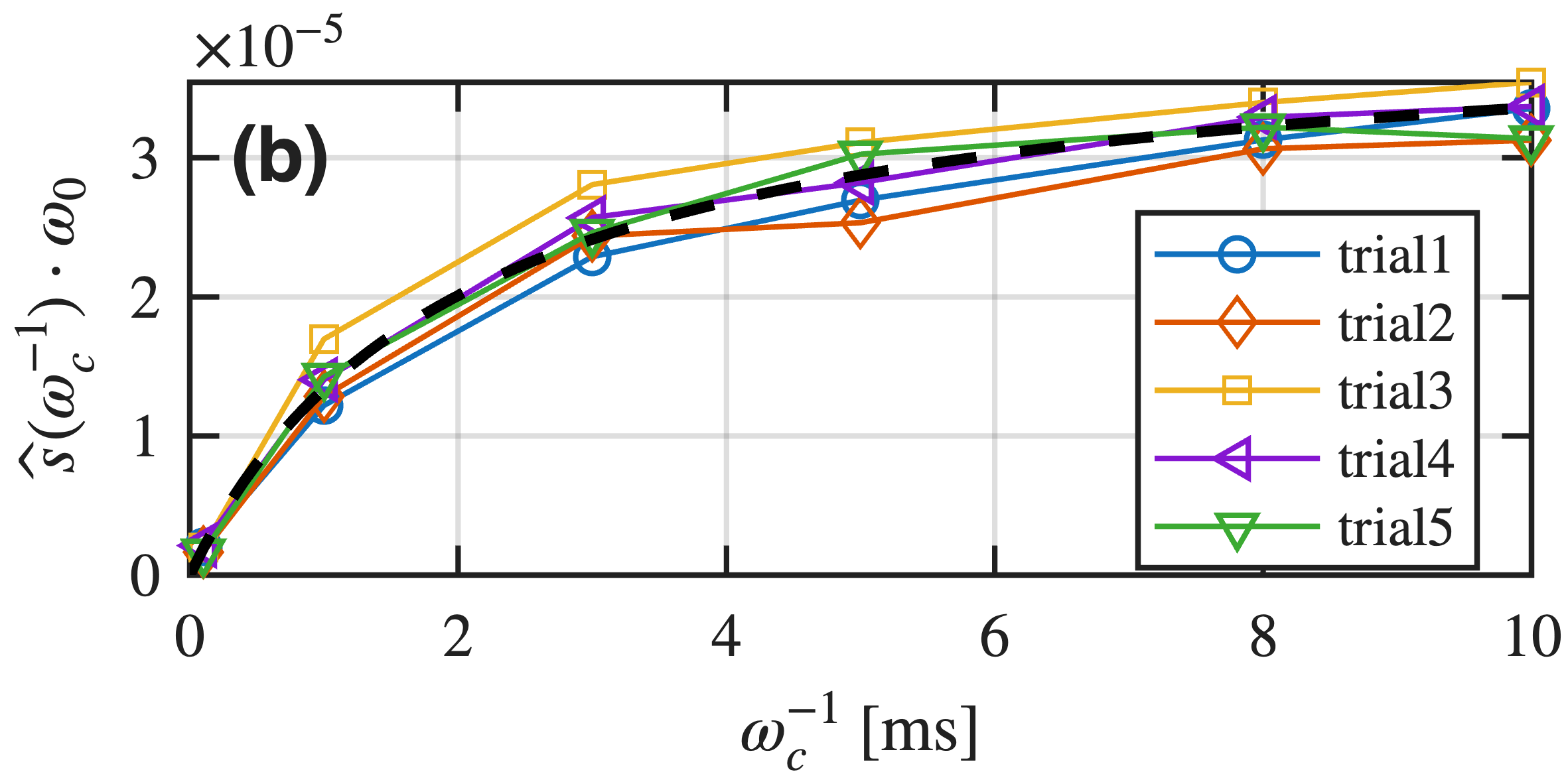}
    
    \includegraphics[width=0.37\textwidth]{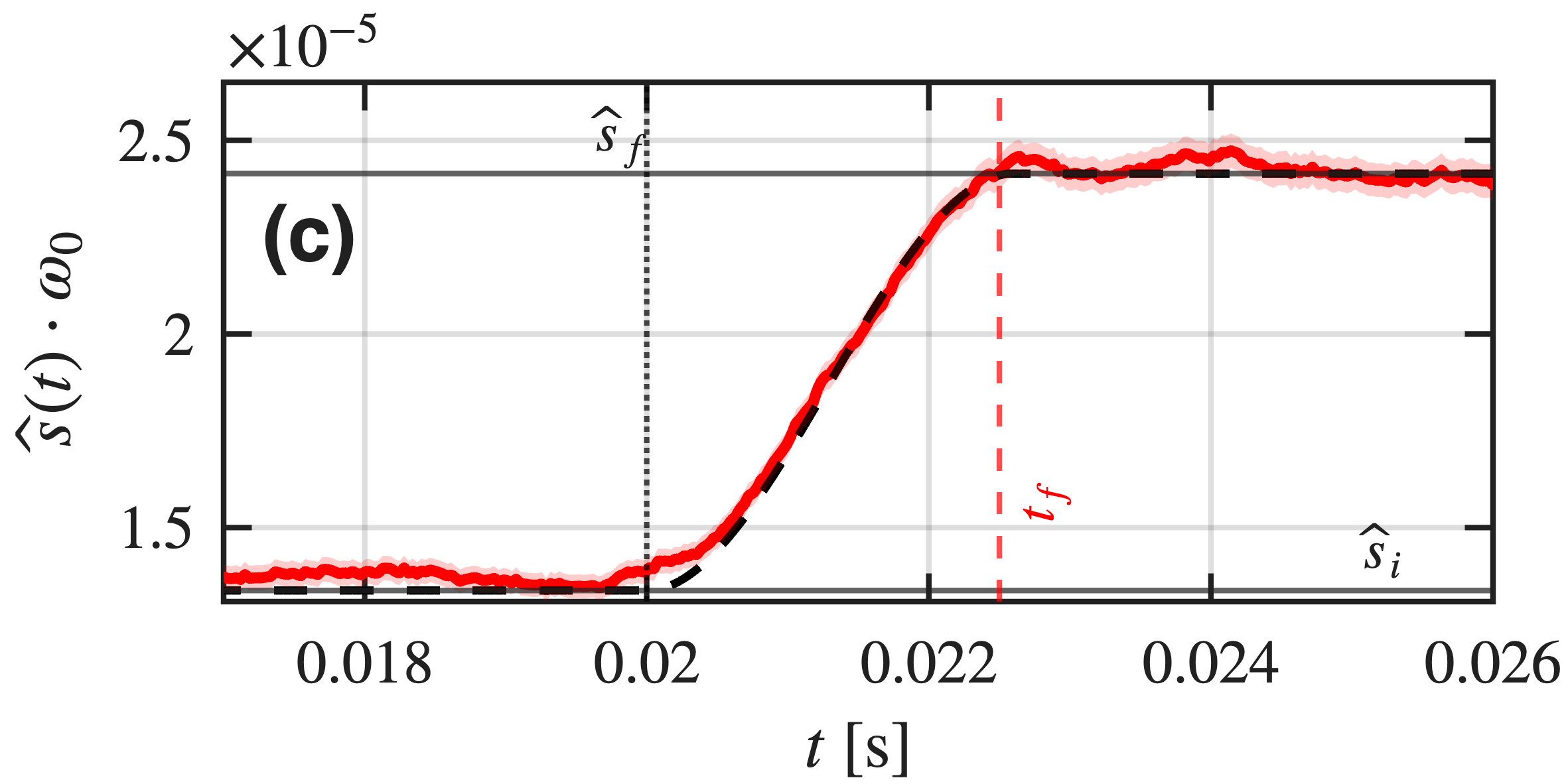}
    \includegraphics[width=0.37\textwidth]{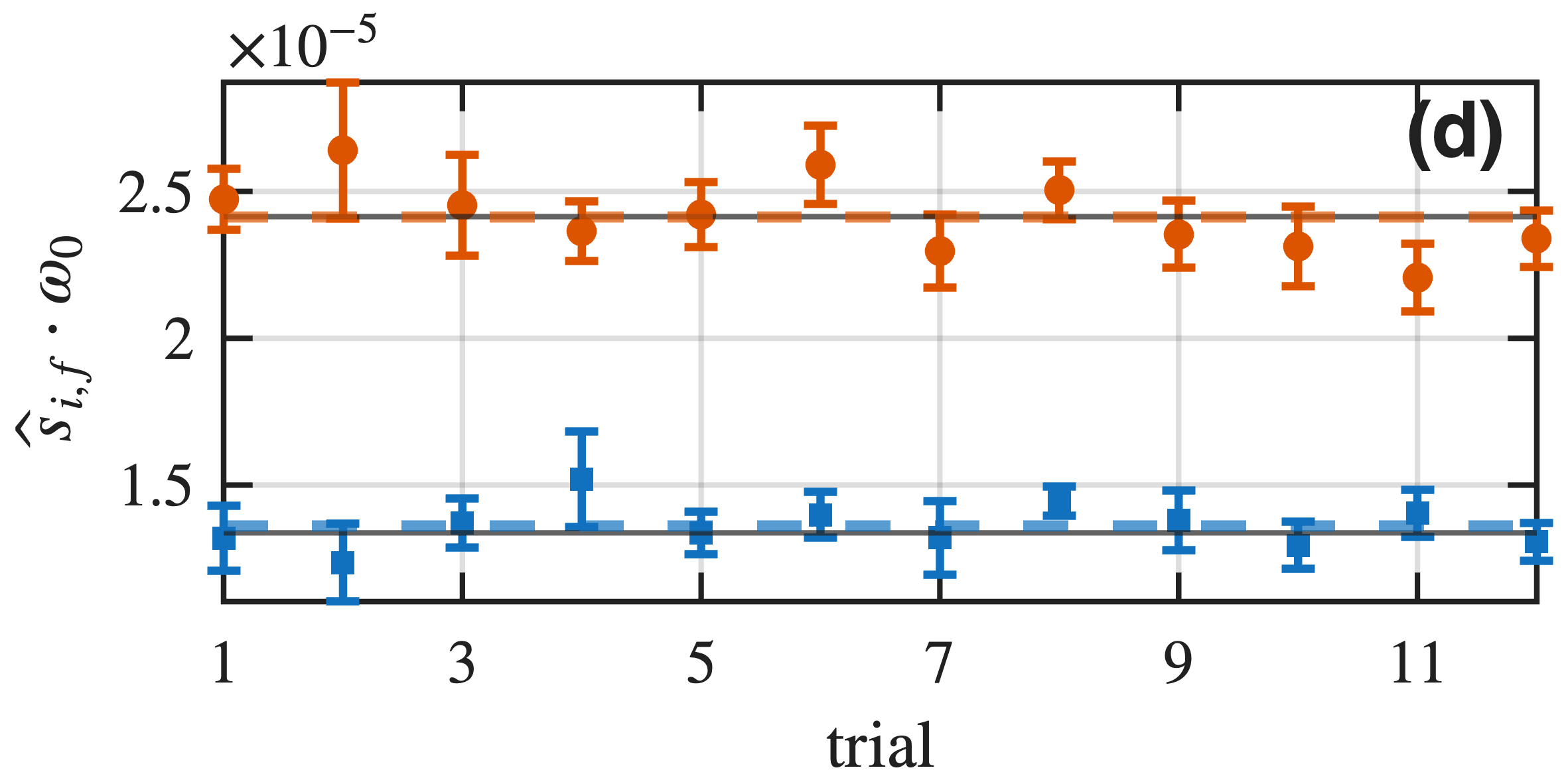}
    
    \includegraphics[width=0.38\textwidth]{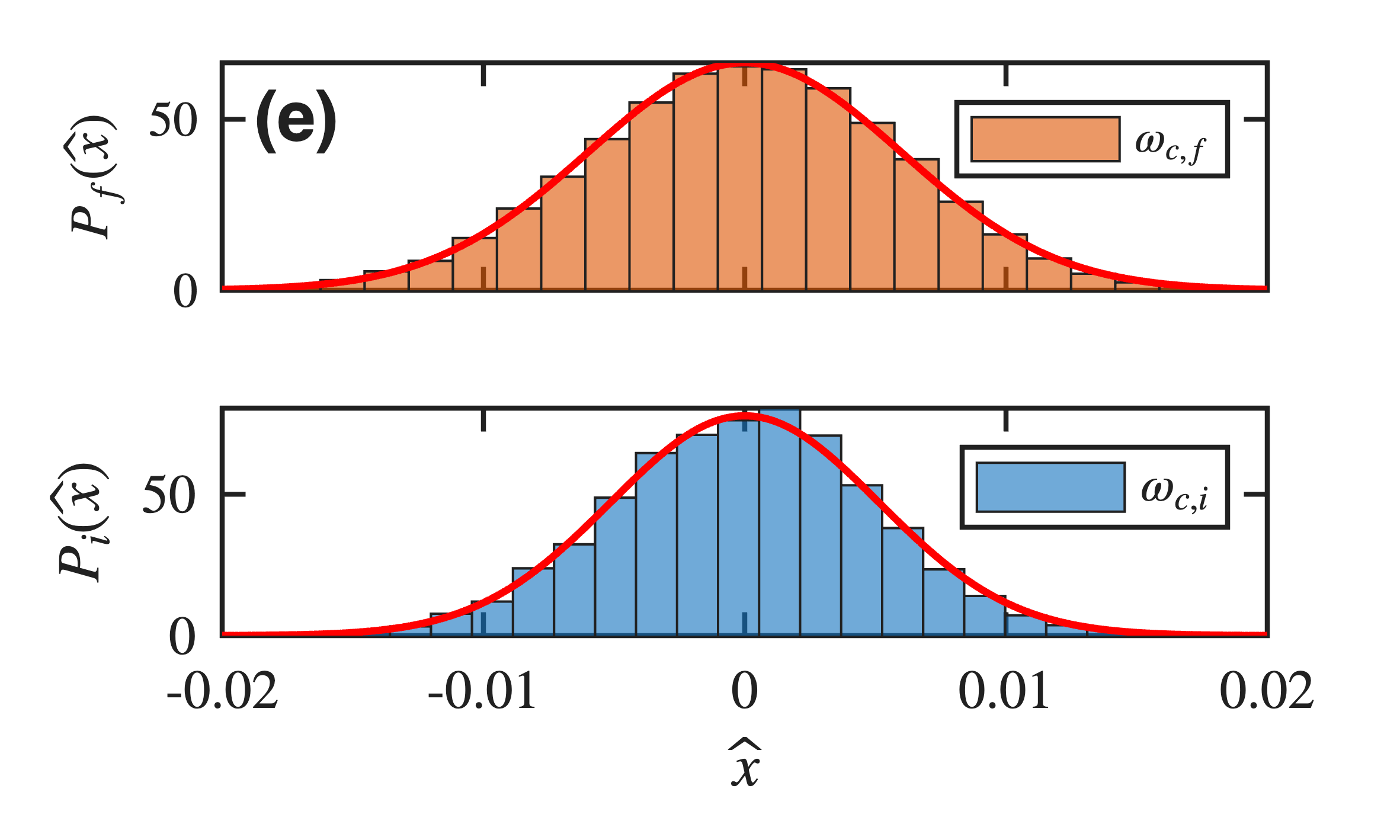}
    \includegraphics[width=0.38\textwidth]{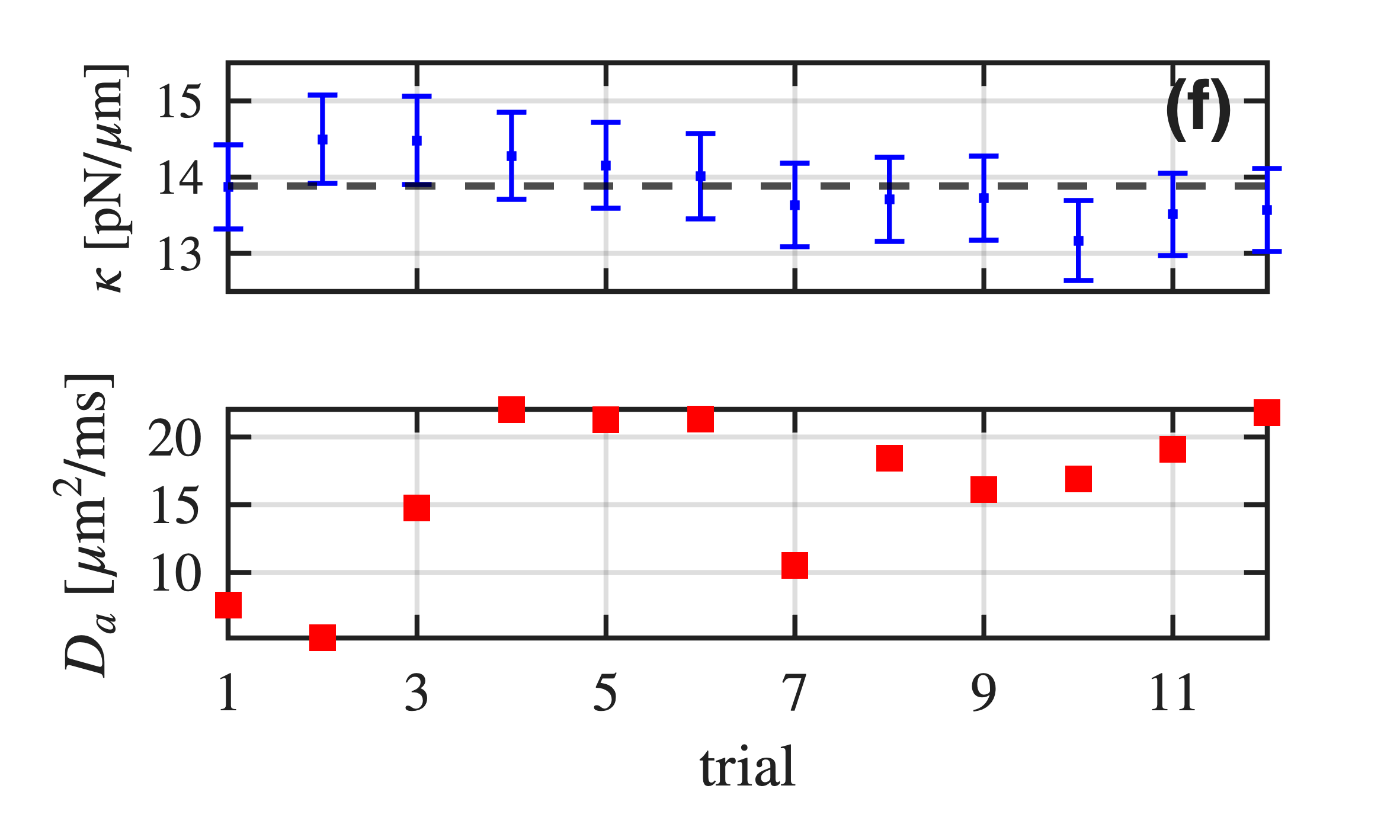}
    \captionsetup{justification=raggedright,singlelinecheck=false}
    \caption{
    \textbf{Experimental calibration and data-acquisition:} \ 
    \textbf{(a)} Stationary variance, $s(\omega_c;D_a)$, as a function of the correlation time $\omega_c^{-1}$ with the active coefficient $D_a$ as a fit parameter.
    \textbf{(b)} Normalized variances $\hat{s}$ of different experimental trials (with different $D_a$ values) collapse onto a universal curve.
    \textbf{(c)} Ensemble-averaged normalized variance $\hat{s}(t)$ during a representative NESS-to-NESS transition (3rd order protocol, red solid curve). The protocol is applied at $t_0$ and reaches the target NESS at the prescribed transition time $t_f=2.60$ms; the black dashed curve shows the analytical prediction.
    \textbf{(d)} Position distributions at the final and initial times of the protocol, together with the corresponding stationary Gaussian distributions (red curves).
    \textbf{(e)} Initial (blue) and final (orange) stationary values $\hat{s}_{i,f}$ measured over successive experimental trials; dashed lines denote the analytical NESS predictions. Error bars indicate standard errors.
    \textbf{(f)} Corresponding trap stiffness $\kappa$ and calibrated active coupling strength $D_a$ for each trial. 
    Experimental (fixed) parameters are: $D=1.41\times10^{-4}~\mu\mathrm{m}^2/\mathrm{ms}$, $\omega_0=0.49~\mathrm{ms}^{-1}$, $\alpha=0.01~\mathrm{Hz}$, and $\omega_c$ is varied between $1/3$ and $1~\mathrm{ms}^{-1}$. 
    }
    \label{fig:exp_SM2}
\end{figure}

\subsection{Data acquisition}

For each engineered active bath protocol $\langle \hat{x}^2\left(t;\omega_c(t)\right) \rangle$, we obtain an ensemble of $M\approx10^4$ particle trajectories. 
The particle motion is detected using a third laser beam ($\lambda=639~\mathrm{nm}$, not shown in Fig.~\ref{fig:exp_SM1}) whose scattered intensity fluctuations are collected and measured by a photo-diode (Thorlabs, Det100A/M). 
The resulting voltage signal is converted into particle position through a calibration of the power spectral density (PSD), measured in equilibrium conditions and fitted to the analytical spectrum of an overdamped particle in a harmonic trap:
\begin{align}
    \textrm{PSD}(f) \propto \frac{D_V}{f^2+f_c^2},
\end{align}
where $\beta_{\textrm{PSD}}\equiv\sqrt{D/D_V}$ is a volt-to-meter (fit) parameter \cite{Berg2004}, and the corner frequency $f_c$ determines the trap stiffness $\kappa=2\pi\gamma f_c$ (cf. Fig.~\ref{fig:exp_SM1}\textbf{(b)}).

Each trajectory is recorded over a time window $T_w=20~\mathrm{ms}$ (30~kHz sampling rate), with the protocol initiated at $t_0=10~\mathrm{ms}$. 
The time window is chosen such that the particle evolves for $>5\omega_c^{-1}$ before and after the protocol, to ensures that the system is well relaxed to the initial NESS before the protocol and to the final NESS by the end of the observation window.

The active coupling strength $D_a$ is calibrated independently before each experimental trial. 
For this purpose, we measure the stationary particle variance $s(\omega_c;D_a)$ for several values of the noise correlation rate $\omega_c$ and fit the resulting dependence to the analytical AOUP steady-state solution (Eq.~(6) in the main text), using $D_a$ as the only fitting parameter (see Fig.~\ref{fig:exp_SM2}\textbf{(a)}). 
We find that the value of $D_a$ is particularly sensitive to experimental conditions and may vary between trials due to small changes in the optical alignment, calibration, or overall preparation of the setup.

To account for these trial-to-trial variations, we analyze the dynamics in terms a normalized position observable, $\hat{x}(t)\equiv x(t)/\sqrt{2D_a}$, and corresponding second moments, including the variance
\begin{align}
    \hat{s}(t)=\frac{\langle x^2(t)\rangle - D/\omega_0 }{2D_a},
\end{align}
and the position-noise covariance
\begin{align}
    \hat{z}(t) = \langle \hat{x}(t)\eta(t)\rangle .
\end{align}
This normalization removes the dependence on $D_a$, allowing measurements obtained from different experimental trials to be compared directly (see Fig.~\ref{fig:exp_SM2}\textbf{(b)}).

The complete procedure is illustrated in Figs.~\ref{fig:exp_SM2}\textbf{(c-f)}. The ensemble-averaged normalized variance $\hat{s}(t)$ resolves the transition between the initial and final NESSs, while the corresponding position distributions at the beginning and end of the protocol recover the respective stationary distributions. Although the calibrated value of $D_a$ varies between experimental trials, the normalized stationary variances remain consistent with the analytical NESS predictions. As an additional stability check, the trap stiffness is measured throughout the experiments and remains approximately constant, confirming that the observed changes in $\hat{s}(t)$ originate from the modulation of the active-noise correlations rather than variations of the confining potential.

\section{Generalized OU Noise Extension}
\label{SM:generalized}

\begin{figure}[t]
    \centering
    \includegraphics[width=\textwidth]{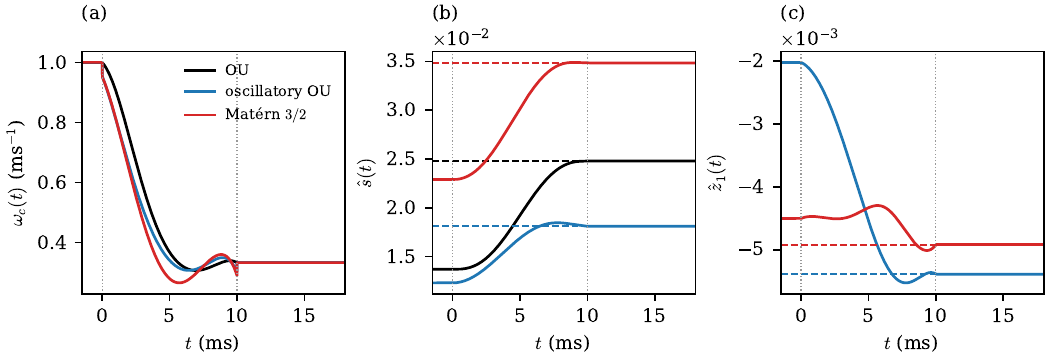}
    \captionsetup{justification=raggedright,singlelinecheck=false}
    \caption{
    \textbf{Reverse engineering for the noises of Table~\ref{tab:S_noises}:}
    expansion $\omega_c:1\to1/3~\mathrm{ms}^{-1}$ with $t_\ff=10~\mathrm{ms}$ for the Ornstein--Uhlenbeck (black), oscillatory OU (blue), and Mat\'ern $3/2$ (red) noises, with $\omega_0=0.49~\mathrm{ms}^{-1}$, $\alpha=0.01$, $w=0.5~\mathrm{ms}^{-1}$, and $\omega_c=\omega_{c,\ff}$ for $t>t_\ff$.
    \textbf{(a)} Reconstructed protocols $\omega_c(t)$.
    \textbf{(b)} Variance $\hat s(t)$.
    \textbf{(c)} Hidden covariance $\hat z_1(t)$.
    Dashed lines mark the values in the target NESS.
    }
    \label{fig:SM_generalized}
\end{figure}

The reverse-engineering approach of the main text relies only on the fact that the second moments of $(x,\eta)$ obey a closed set of equations under the control $\omega_c(t)$. This remains true when the active noise is the first component of a linear system driven by white noise,
\begin{equation}
\eta=c^{\mathsf T}y,
\qquad
dy=\bigl[\omega_c(t)A_0+A_1\bigr]y\,dt+\sqrt{\omega_c(t)}\,B_0\,dW,
\label{eq:S_embedding}
\end{equation}
with $c=(1,0,\dots,0)^{\mathsf T}$ and $y\in\mathbb R^m$. The matrices are chosen such that the stationary covariance $S$ of $y$ does not depend on $\omega_c$, so that, as for $\langle\eta^2\rangle=\alpha$ in the main text, the protocol changes only the rate at which the noise decorrelates and not its one-time statistics. Table~\ref{tab:S_noises} lists the noises considered here: the Ornstein--Uhlenbeck noise of the main text and two noises with $m=2$, whose correlation functions oscillate or are smoother. Figure~\ref{fig:SM_generalized} shows the transitions constructed below.

\begin{table}[h!]
    \centering
    \renewcommand{\arraystretch}{1.3}
    \setlength{\tabcolsep}{8pt}
    \begin{tabular}{l c c c c}
        \hline
        Noise & $m$ & $A_0$ & $A_1$ & $\langle\eta(t)\eta(t+\tau)\rangle$ at fixed $\omega_c$ \\
        \hline
        Ornstein--Uhlenbeck & $1$ & $-1$ & $0$ & $\alpha\,e^{-\omega_c|\tau|}$ \\
        oscillatory OU & $2$ & $-I$ & $w\!\m{0&1\\-1&0}$ & $\alpha\,e^{-\omega_c|\tau|}\cos(w\tau)$ \\
        Mat\'ern $3/2$ & $2$ & $\m{0&1\\-1&-2}$ & $0$ & $\alpha\,(1+\omega_c|\tau|)\,e^{-\omega_c|\tau|}$ \\
        \hline
    \end{tabular}
    \caption{Active noises of the form~\eqref{eq:S_embedding} and their correlation functions at fixed $\omega_c$. For all three, the stationary covariance is $S=\alpha I$ and $B_0B_0^{\mathsf T}=-(A_0S+SA_0^{\mathsf T})$.}
    \label{tab:S_noises}
\end{table}

With $\hat{\mathbf z}=\langle x\,y\rangle/\sqrt{2D_a}$, whose first component $\hat z_0$ is the normalized $\langle x\eta\rangle$ of the main text, the second moments obey
\begin{equation}
\dot{\hat s}=-2\omega_0\hat s+2\hat z_0,
\qquad
\dot{\hat{\mathbf z}}=\bigl(\omega_cA_0+A_1-\omega_0I\bigr)\hat{\mathbf z}+\alpha c,
\label{eq:S_gen_moments}
\end{equation}
where the source term $\alpha c=Sc$ uses that the covariance of $y$ remains equal to $S$ under any protocol. For $m=1$, Eq.~\eqref{eq:S_gen_moments} reduces to Eq.~\eqref{eq:S_moments}.

As in the main text, the prescribed variance fixes $\hat z_0=\omega_0\hat s+\tfrac12\dot{\hat s}$, and the first row of the equation for $\hat{\mathbf z}$ can be solved for $\omega_c$. For $m=2$, however, this row also contains the hidden covariance $\hat z_1$, which the prescribed variance does not determine, so that $\omega_c$ and $\hat z_1$ must be found together from both rows:
\begin{align}
\text{oscillatory OU:}&\quad
\omega_c=\frac{\alpha-\dot{\hat z}_0+w\hat z_1}{\hat z_0}-\omega_0,
\qquad
\dot{\hat z}_1=-(\omega_0+\omega_c)\hat z_1-w\hat z_0,
\nonumber\\
\text{Mat\'ern }3/2:&\quad
\omega_c=\frac{\dot{\hat z}_0+\omega_0\hat z_0-\alpha}{\hat z_1},
\qquad
\dot{\hat z}_1=-\omega_c(\hat z_0+2\hat z_1)-\omega_0\hat z_1 .
\label{eq:S_compatibility}
\end{align}
Substituting $\omega_c$ into the equation for $\dot{\hat z}_1$ gives the compatibility condition, a closed ordinary differential equation for $\hat z_1$ driven by the prescribed variance. We integrate it forward from the value of $\hat z_1$ in the initial NESS, and $\omega_c(t)$ then follows from the first relation in Eq.~\eqref{eq:S_compatibility}, so that $\hat s$ follows the prescribed trajectory exactly on $[0,t_\ff]$.

This changes how the target NESS is reached at $t_\ff$. In the main text, $\hat z_0$ is an algebraic function of $\hat s$ and $\dot{\hat s}$, so the boundary conditions on the endpoints and derivatives of $\hat s$ fix $\hat z_0$ at $t_\ff$. The hidden covariance has no such expression: $\hat z_1(t_\ff)$ is the endpoint of the compatibility condition and depends on the whole prescribed trajectory, so conditions on the endpoints and differentiability of $\hat s$ alone cannot fix it, and it generally differs from its target $\hat z_{1,\rm ss}(\omega_{c,\ff})$. The system then keeps evolving after $t_\ff$ under $\omega_{c,\ff}$; with $p_5^{\rm cts}$, the variance of the oscillatory noise leaves $\hat s_\ff$ by up to about $6\times10^{-3}(\hat s_\ff-\hat s_\ii)$ before returning to it. We therefore impose $\hat z_1(t_\ff)=\hat z_{1,\rm ss}(\omega_{c,\ff})$ as an additional terminal condition and meet it with a free coefficient of the fifth-order class: keeping $a_2=3$, as in $p_5^{\rm cts}$, we solve the compatibility condition forward for a given $a_1$ and adjust $a_1$ until $\hat z_1(t_\ff)$ reaches its target. Each hidden covariance adds one such terminal condition, and hence one coefficient to be tuned.

Figure~\ref{fig:SM_generalized} applies this construction to the expansion $\omega_c:1\to1/3~\mathrm{ms}^{-1}$ in $t_\ff=10~\mathrm{ms}$. The terminal condition gives $(a_1,a_2)=(4.670,3)$ for the oscillatory noise and $(1.794,3)$ for the Mat\'ern $3/2$ noise, while the Ornstein--Uhlenbeck noise uses $p_5^{\rm cts}$. For all three noises, $\hat s$ reaches its target at $t_\ff$, as does $\hat z_1$ for the two noises with $m=2$, and both remain there afterwards [Figs.~\ref{fig:SM_generalized}(b,c)]. Since $p''=2a_1\neq0$ at both ends of the tuned shapes, the corresponding protocols jump at both boundaries [Fig.~\ref{fig:SM_generalized}(a)], and their variance slightly overshoots its target before $t_\ff$.

\end{document}